\documentclass[a4paper,fleqn,usenatbib]{mnras}

\usepackage{newtxtext,newtxmath}

\usepackage[T1]{fontenc}
\usepackage{ae,aecompl}

\usepackage{graphicx}	
\usepackage{amsmath}	
\usepackage{amssymb}	

\title[Spatially-Dependent Modelling of PWN G0.9+0.1]{Spatially-Dependent Modelling of Pulsar Wind Nebula G0.9+0.1}

\author[C. van Rensburg et al.]{
C. van Rensburg,$^{1}$\thanks{E-mail: carlo.rensburg@gmail.com}
P.P. Kr\"uger$^{1}$
and C. Venter$^{1}$
\\
$^{1}$Centre for Space Research, North-West University, Potchefstroom Campus, Private Bag X6001, Potchefstroom, South Africa, 2520\\
}

\date{Accepted 2018 March 19. Received 2018 March 19; in original form 2017 March 3}

\pubyear{2017}

\begin{document}
\label{firstpage}
\pagerange{\pageref{firstpage}--\pageref{lastpage}}
\maketitle

\begin{abstract}
We present results from a leptonic emission code that models the spectral energy distribution of a pulsar wind nebula by solving a Fokker-Planck-type transport equation and calculating inverse Compton and synchrotron emissivities. We have created this time-dependent, multi-zone model to investigate changes in the particle spectrum as they traverse the pulsar wind nebula, by considering a time and spatially-dependent $B$-field, spatially-dependent bulk particle speed implying convection and adiabatic losses, diffusion, as well as radiative losses. Our code predicts the radiation spectrum at different positions in the nebula, yielding the surface brightness versus radius and the nebular size as function of energy. We compare our new model against more basic models using the observed spectrum of pulsar wind nebula G0.9+0.1, incorporating data from H.E.S.S.\ as well as radio and X-ray experiments. We show that simultaneously fitting the spectral energy distribution and the energy-dependent source size leads to more stringent constraints on several model parameters.
\end{abstract}




\section{Introduction}
\label{sec:intro}
Pulsar wind nebulae (PWNe) are true multi-wavelength objects, observable from the highest $\gamma$-ray energies down to the radio waveband, sometimes exhibiting complex morphologies in the different energy domains. Discoveries during the last decade by ground-based Imaging Atmospheric Cherenkov Telescopes (IACTs) have increased the number of known very-high-energy (VHE, $E >$ 100 GeV) $\gamma$-ray sources to nearly 200\footnote{http://tevcat2.uchicago.edu/}. \cite{Hewitt2015} note that nearly 40 of these are confirmed pulsar wind nebulae (PWNe). Following the nine-year H.E.S.S. Galactic Plane Survey (HGPS; \citealt{HGPS2016}), H.E.S.S.\ published a paper describing the properties of 19 PWNe and 10 strong PWN candidates, as well as empirical trends between several PWN/pulsar parameters \citep{PWN_pop2017}. It is expected that the future Cherenkov Telescope Array (CTA), with its order-of-magnitude increase in sensitivity and improvement in angular resolution, will discover several more (older and fainter) PWNe and reveal many more morphological details. A systematic search with the \textit{Fermi} Large Area Telescope (LAT) for GeV emission in the vicinity of TeV-detected sources yielded 5 high-energy $\gamma$-ray PWNe and 11 PWN candidates \citep{3FGL2015}. In the X-ray to VHE $\gamma$-ray energy range there are 85 PWNe or PWN candidates with 71 of them having associated pulsars \citep{Kargaltsev2012}. 

For slower moving pulsars, one might observe a composite supernova remnant (SNR), with nebular and shell emission visible in both radio and X-ray bands. Such young systems (having ages of a few thousand years) exhibit a high degree of spherical symmetry and it is possible that the SNR reverse shock has not yet interacted with the PWN (e.g., SNR G11.2$-$0.3 and G21.5$-$0.9). The PWN around PSR B1509$-$58 provides a counter example, exhibiting a strong anti-correlation between the radio and X-ray emission morphology. This system is reminiscent of older PWNe associated with fast-moving pulsars and $\gamma$-ray sources that exhibit complex morphologies (e.g., the Rabbit Nebula and G327.1$-$1.1; \citealt{Roberts2005, SlanePWN2017}). In even older PWNe (with ages of tens of thousands of years), a rapidly decreasing $B$-field may lead to $\gamma$-ray emission dominating the observed radio and X-ray emission (e.g., HESS J1825$-$137; \citealt{SlanePWN2017}). 

High-resolution observations by \textit{Chandra} X-ray Observatory have furthermore revealed complex substructures such as toroidal structures, bipolar jets, and filaments \citep{Helfand2001, Roberts2003}. Similarly, high-resolution radio images sometimes reveal complex PWN morphology including filaments, knots, and holes \citep{Dubner2008}. Complementary optical and infrared observations may uncover spectral features in the particle spectrum, information about the shocked supernova ejecta, and newly formed dust \citep{Slane2017_T}.

PWNe (plerions) have historically been identified based on their observational properties, i.e., having a filled-centre emission morphology, a flat spectrum at radio wavelengths, and a very broad spectrum of non-thermal emission ranging from the radio band to high-energy $\gamma$-rays \citep[e.g.,][]{Weiler1978, deJager_2009, Amato2014}. Apart from the Galactic population of PWNe, H.E.S.S. has detected a powerful extragalactic PWN in the Large Magellanic Cloud (LMC) lying at a distance of $\sim$50~kpc \citep{Abramowski2012}. Galactic PWNe are interesting laboratories due to the fact that they are nearby sources that are well resolved, especially in the X-ray band. The knowledge that we gain from studying them also has a strong impact in many other fields, ranging from $\gamma$-ray bursts (GRBs) to active galactic nuclei (AGNs).

Current spectral models (mostly leptonic) attempt to reproduce the observed spectral energy distributions (SEDs) of PWNe.  These models, however, differ slightly from one another. Most of them model the structure of the PWN as a single sphere (i.e., one-zone models) by assuming spherical symmetry, see e.g.,  \cite{VdeJager2007}, \cite{Zhang_2008}, \cite{Tanaka_Takahara_2011}, \cite{Martin2012} and \cite{Torres2014}, although time dependence is an important feature of these models. The majority of models assume that the injection spectrum of particles is in the form of a broken power law. Using a particle-in-cell code, \cite{Sironi2011} found that the particle injection spectrum may be a modified Maxwellian with a power-law tail with index $\sim$ 1.5. This is the result of magnetic reconnection at the termination shock due to the striped wind of the PWN. \cite{Vorster2013}, on the other hand, model the injection of particles as a type of broken power-law, with the exception that the flux at the break energy may vary discontinuously (i.e., assuming a multi-component injection spectrum). Some authors model the injection of particles including acceleration due to the SNR shock and their subsequent emission. This is also known as thermal leakage, see \cite{Fang2010}.

The particle transport is also handled differently. Some works calculate the particle population solving a differential equation involving diffusion, convection, and adiabatic and radiation losses \citep[e.g.,][]{Tanaka_Takahara_2011, Torres2014}, while other models only consider particle escape \citep[e.g.,][]{Zhang_2008, Qiao_2009}. Some models omit adiabatic losses \citep[e.g.,][]{VdeJager2007}, and there are different specifications for the time-dependent PWN $B$-field. Different models also consider different types of emission; for example, \cite{Tanaka_Takahara_2011} take synchrotron-self-Compton emission into account and others consider bremsstrahlung \citep[e.g.,][]{Martin2012} in addition to inverse Compton (IC) scattering and synchrotron radiation (SR). While these models have been reasonably successful at reproducing observed SEDs, these one-zone spectral models can not reproduce any of the observed morphological properties of PWNe. 

Conversely, magnetohydrodynamic (MHD) models are also being developed that can model the morphology of PWNe in great detail \citep[e.g.,][]{Bucciantini2014}, but they in turn cannot predict the SED from the PWNe. These models describe the geometry and environment of the PWN and not the high-energy particle spectrum and therefore the information about the radiation spectrum is lost. There are, however, models that follow a hybrid approach \citep[e.g.,][]{Porth2016}: they model the morphology of the PWN in great detail using an MHD code, and then use a steady-state spectral model to produce the SED of the PWN. 

In light of the above, there is a void in the current modeling landscape for a spatio-temporal and energy-dependent PWN model that models both morphology and the SED of a PWN. By adding a spatial dimension to an emission code, one is able to constrain the model more significantly using available data such as surface brightness profiles, spectral index versus radius, and energy-dependent source size, and thus probe the PWN physics more deeply. For example, current spectral models have degenerate best-fit parameters. Since they are not spatially dependent, they cannot constrain functional dependencies such as $B$-field and diffusion coefficient profiles. Our newly developed time-dependent, multi-zone model aids in breaking some degeneracies by first constraining the profiles of, e.g., the PWN $B$-field and then fitting the observed SED in a more constrained parameter space, thus making use of both spectral and spatial data. Further motivation to develop this type of model comes from observations of a PWN population. \citet{Kargaltsev2015} found that the measured $\gamma$-ray luminosity (1$-$10 TeV) of PWNe does not correlate with the spin-down luminosity of their embedded pulsars. Alternatively, they found that the $X$-ray luminosity (0.5$-$8 keV) is correlated with the pulsar spin-down luminosity. Furthermore, there are indications that a strong correlation exists between the TeV surface brightness of the PWNe and the spin-down luminosity of their embedded pulsars \citep{PWN_pop2017}. A spatially-dependent spectral model will yield the flux as a function of radius, allowing one to model the surface brightness and thus probe this and other relationships. Furthermore, one would be in a position to interpret the anticipated morphological details that will be measured by future experiments. In light of the above, we implemented such a model \citep{vRensburg2014, CvR2015} and discuss its behaviour in this paper\footnote{It has come to our attention during the final stages of our code development that an independent study has recently been published by \cite{Lu2017} describing a spatially-dependent PWN model. Their code models the electron spectrum using a Fokker-Planck equation, similar to our code, and then predicts the SED from the PWN. They also assume spherical symmetry and take into account SR, IC, and adiabatic energy losses. They use free expansion to determine the radius of the PWN, while we in contrast use the surface brightness of the PWN to predict its size. Although they calculate the surface brightness, they only predict this for the X-ray band. We model the surface brightness for the entire electromagnetic spectrum and thus we predict the size of the PWN as a function of energy. We also perform a thorough parameter study to show the effects of all the model parameters. Lastly, they applied their model to MSH 15-52 while we studied G0.9+0.1. We therefore believe that our results are complementary to their work. Indeed, their work provides another independent calibration of our model, and we find very similar spectra for the same input parameters for MSH 15-52. See Section~\ref{sec:cal} for other calibrations of our code.}.

In this paper, we describe the development of our PWN model.  In Section~\ref{sec:model}, we describe some technical details and assumptions of our model. Section~\ref{sec:cal} details the calibration of our code against independent models, while we perform a parameter study in Section~\ref{sec:Par}. We discuss spatially-dependent results in Section \ref{sec:Space}, and our conclusions follow in Section~\ref{sec:concl}. 

\section{The Model}
\label{sec:model}
In this section the development and implementation of the multi-zone, time-dependent code, which models the transport of particles through a PWN, is described. We make the simplifying assumption that the geometrical structure of the PWN may be modelled as a sphere into which particles are injected and allowed to diffuse and undergo energy losses. Another assumption is that particle transport is spherically symmetric and thus the only changes in the particle spectrum will be in the radial direction (apart from changes in the particle energy). The model therefore consists of three dimensions in which the transport equation is solved: the spatial or radial dimension, the lepton energy dimension, and the time dimension. 

\subsection{The Transport Equation}
We solve a Fokker-Planck-type equation that includes diffusion, convection, energy losses (radiative and adiabatic), as well as a particle source term. We start from the following form of the transport equation \citep{Moraal2013}
\begin{equation}
\frac{\partial f}{\partial t} = -\nabla \cdot \mathbf{S} + \frac{1}{p^2}\frac{\partial }{\partial p}\left(p^2\left\langle\dot{p}\right\rangle_{\rm{tot}} f\right) + Q(\mathbf{r},\mathbf{p},t),
\label{kopp1}
\end{equation}
with $f$ the distribution function (number of particles per six-dimensional unit phase-space volume, spanning three spatial and three momentum directions), $Q(\mathbf{r},\mathbf{p},t)$ the particle injection spectrum, $r$ the radial dimension, $p$ the particle momentum, and $\langle\dot{p}\rangle _{\rm{tot}}$ the total rate of change of $p$. The term $\nabla \cdot \mathbf{S} = \nabla \cdot \left(\mathbf{V}f - \mathbf{\underline{K}} \nabla f\right)$ describes the general movement of particles in the PWN, with $\mathbf{V}$ the bulk motion of particles, $\mathbf{\underline{K}}$ the diffusion tensor, and $\mathbf{S}$ the streaming density. However, we rewrite Eq.~(\ref{kopp1}) in terms of energy and also transform the distribution function to a particle spectrum per unit volume $U_{\rm{p}}$ as is more customary. Following \cite{Moraal2013}, we use the relation $U_{\rm{p}}(\mathbf{r},p,t) = 4\pi p^2 f(\mathbf{r},\mathbf{p},t)$ (with the units of $U_{\rm{p}}$ being number of particles per three-dimensional volume per momentum interval) to convert $f$ to a particle spectrum, and $E^2 = p^2c^2 + E_0^2$ to convert Eq.~\eqref{kopp1} from momentum to energy space, with $E_0 = m_{\rm{e}}c^2$, $m_{\rm{e}}$ the electron mass, and $c$ the speed of light in vacuum. We also assume that the diffusion is only energy dependent, $\mathbf{\underline{K}} = \kappa(E_{\rm{e}})$, with $E_{\rm{e}}$ the lepton energy. Thus Eq.~\eqref{kopp1} becomes
\begin{equation}
\begin{split}
\frac{\partial N_{\rm{e}}}{\partial t} =& -\mathbf{V} \cdot (\nabla N_{\rm{e}}) +  \kappa \nabla^2 N_{\rm{e}}  \\
&+ \frac{1}{3}(\nabla \cdot \mathbf{V})\left( \left[\frac{\partial N_{\rm{e}}}{\partial \ln E_{\rm{e}}} \right] - 2N_{\rm{e}} \right)   \\
&+ \frac{\partial }{\partial E}(\dot{E}_{\rm{e,tot}}N_{\rm{e}}) +  Q(\mathbf{r},E_{\rm{e}},t),
\end{split}
\label{eq:transportFIN}
\end{equation} 
with $\dot{E}_{\rm{e,tot}}$ total energy loss rate, including radiation and adiabatic energy losses. The units of $N_{\rm{e}} \equiv U_{\rm{E}}(\mathbf{r},E_{\rm{e}},t)$ are the number of particles per unit energy and volume.  See \cite{CvR2015} for more details. 

\subsection{The Particle Injection Spectrum}\label{sec:Injec}
Following \cite{VdeJager2007}, we use a broken power law for the particle injection spectrum
\begin{equation}
Q(E_{\rm{e}},t) = \left\{\begin{matrix}
Q_0(t)\left(\frac{E_{\rm{e}}}{E_{\rm{b}}}\right)^{\alpha_1} \qquad E_{\rm{e,min}} \leq E_{\rm{e}}<E_{\rm{b}}\\ 
Q_0(t)\left(\frac{E_{\rm{e}}}{E_{\rm{b}}}\right)^{\alpha_2} \qquad E_{\rm{b}} < E_{\rm{e}} \leq E_{\rm{e,max}}.
\end{matrix}\right.
\label{brokenpowerlaw}
\end{equation}  
Here $Q_0(t)$ is the time-dependent normalisation constant, $E_{\rm{b}}$ the break energy, $\alpha_1$ and $\alpha_2$ are the spectral indices. To obtain $Q_0$ we use the following form for the spin-down luminosity of the pulsar: $L(t) = L_0/\left(1+t/\tau_0\right)^2$ assuming a braking index of $n=3$ \citep[e.g.,][]{Reynolds1984}. The birth characteristic age is $\tau_{0} = P_0/(n-1)\dot{P}_0$, $t$ is the time, $L_0$ the initial spin-down luminosity, and $P_0$ and $\dot{P_0}$ are the pulsar's initial period and time derivative of the period. From the current value of $P$ and $\dot{P}$, we first calculate $\tau_{\rm{c}} = P/(n-1)\dot{P}$, which is the characteristic age of the pulsar \citep{Gaensler06}. Next we use $\tau_0 = \tau_{\rm{c}} - t_{\rm{age}}$, with $t_{\rm{age}}$ the age of the PWN. From this follows $L(t)$ for constant $n$, with $t_{\rm{age}}$ the only free parameter (see Appendix). To solve for $Q_0$ we set
\begin{equation}
  \epsilon L(t) = \int_{E_{\rm{min}}}^{E_{\rm{b}}}QE_{\rm{e}}dE_{\rm{e}} + \int_{E_{\rm{b}}}^{E_{\rm{max}}}QE_{\rm{e}}dE_{\rm{e}},
  \label{normQ}
\end{equation}
with $\epsilon = 1/(1+\sigma)$ the constant conversion efficiency of the spin-down luminosity to particle power, and $\sigma$ the ratio of electromagnetic to particle energy density.

\subsection{Energy Losses}
Energy losses in our model are due to two main processes -- radiative and adiabatic energy losses. For radiative energy losses we incorporated synchrotron radiation (SR) and inverse Compton (IC) scattering\footnote{We neglect synchrotron self Compton (SSC) and Bremsstrahlung as the energy losses due to these two effects are orders of magnitude smaller than SR and IC scattering for PWN G0.9+0.1.}, similar to calculations done by \cite{Kopp2013} in their globular cluster model. The SR losses are given by \citep{BlGould1970}
\begin{equation}
  \left(\frac{dE_{\rm{e}}}{dt}\right)_{\rm{SR}} = -\frac{\sigma_T c}{6\pi E_0^2} E_{\rm{e}}^2B_{\bot}^2,
  \label{SR}
\end{equation}
with $\sigma_T = (8\pi/3)r_{\rm{e}}^2 = 6.65 \times 10^{-25}\rm{cm}^2$ the Thomson cross section, $B_{\bot}$ the average perpendicular PWN $B$-field at a certain time and radius, and $r_{\rm{e}} = e^2/m_{\rm{e}}c^2$ the classical electron radius. The IC scattering energy loss rate of leptons scattering blackbody (BB) photons is given by
\begin{equation}
\begin{split}
  \left(\frac{dE_{\rm{e}}}{dt}\right)_{\rm{IC}} = \,\,\,\,\, &  \frac{g_{\rm{IC}}}{E_{\rm{e}}^2} \sum_{l=1}^{3} \int \!\!\!\! \int n_{\varepsilon,l}(r,\varepsilon,T_l)\,\,\, \times \\
  &\frac{E_\gamma}{\varepsilon} \zeta(E_{\rm{e}},E_{\gamma},\varepsilon) d \varepsilon dE_\gamma,
  \end{split}
  \label{IC}
\end{equation}
with $n_{\varepsilon,l}(r,\varepsilon,T_l)$ the BB photon number density of the $l^{th}$ BB component, $g_{\rm{IC}} = 2\pi e^4c$, $\varepsilon$ the soft-photon energy, $T_{l}$ the BB temperature, $E_{\gamma}$ the TeV upscattered photon energy, and $\zeta$ the collision rate
\begin{equation}
\zeta(E_{\rm{e}},E_{\gamma},\varepsilon) = \zeta_0 \hat{\zeta}(E_{\rm{e}},E_{\gamma},\varepsilon),
\end{equation} 
with $\zeta_0 = 2\pi e^4E_0c/\varepsilon E_{\rm{e}}^2$, and $\hat{\zeta}$ given by \citep{Jones1968}
\begin{equation}
\hat{\zeta}(E_{\rm{e}},E_{\gamma},\varepsilon) =\left\{\begin{matrix}
0 &\rm{if} &E_{\gamma} \leq \frac{\varepsilon E_0^2}{4 E_{\rm{e}}^2},\\ 
\frac{E_{\gamma}}{\varepsilon}-\frac{E_0^2}{4 E_{\rm{e}}^2} &\rm{if} &\frac{\varepsilon E_0^2}{4 E_{\rm{e}}^2} \leq E_{\gamma} \leq \varepsilon,\\ 
f(q,g_0) &\rm{if} &\varepsilon \leq E_{\gamma} \leq \frac{4\varepsilon E_{\rm{e}}^2}{E_0^2 + 4\varepsilon E_{\rm{e}}},\\ 
0 &\rm{if} &E_{\gamma} \geq \frac{4\varepsilon E_{\rm{e}}^2}{E_0^2 + 4\varepsilon E_{\rm{e}}}.
\end{matrix}\right.
\end{equation}
Here, $f(q,g_0) = 2q$ln$q+(1-q)(1+(2+g_0)q)$, $q=E_0^2E_{\gamma}/(4\varepsilon E_{\rm{e}}(E_{\rm{e}}-E_{\gamma}))$, and $g_0(\varepsilon,E_{\gamma}) = 2\varepsilon E_{\gamma}/E_0^2$.

The particles in the PWN also lose energy due to adiabatic processes caused by the bulk motion of the particles in the PWN as energy is expended to expand the PWN. The adiabatic energy losses are given by $\dot{E}_{\rm{e,ad}} = \frac{1}{3}(\nabla \cdot \mathbf{V})E_{\rm{e}}$ \citep[e.g.,][]{Zhang_2008}. The two radiation loss rates and the adiabatic energy loss rate can be added to find the total loss rate $\dot{E}_{\rm{e,tot}}$ used in Eq.~(\ref{eq:transportFIN}).

\subsection{Diffusion}
The particle diffusion is assumed to be Bohm-type diffusion, with the scalar diffusion coefficient $\kappa$ given by
\begin{equation}
  \kappa(E_{\rm{e}}) = \kappa_B\frac{E_{\rm{e}}}{B},
  \label{eq:kappa}
\end{equation}
with $\kappa_B = c/3e$, $e$ denoting the elementary charge. We are currently unaware how turbulent the $B$-field is inside the PWN, although we have some constraints from the polarised radio spectrum. Due to this uncertainty we are not sure what form of diffusion coefficient to use and therefore we choose Bohm diffusion as a first approximation when fitting spectral and spatial data. This is a fairly common practice as it describes slow diffusion that is perpendicular to the local $B$-field\footnote{We treat the $B$-field as predominantly azimuthal as is standard practice, e.g., \citet{Schock2010, Vorster2014}. This assumption is based on several arguments: in this case $\nabla \cdot \mathbf{B} = 0$, at typical PWN scales any dipolar field components have all but died out compared to the toroidal components (given their respective $1/r^3$ vs.\ $1/r$ decay), and X-ray observations show ubiquitous polar and equatorial outflows supporting an azimuthal structure winding around the pulsar in the equatorial plane. Lastly, radio polarisation measurements indicate that the magnetic field must be very ordered.}.

In the parameter study we perform in Section~\ref{sec:Par}, however, we parametrise the diffusion coefficient as  
\begin{equation}
\kappa(E_{\rm{e}}) = \kappa_0 \left( \frac{E_{\rm{e}}}{E'_0} \right)^q,
 \label{eq:kappa123}
\end{equation}
with $E'_0 = 1$ TeV (with Bohm diffusion being a special case of this general parametric form). This allowed us to change the normalisation of the diffusion coefficient using $\kappa_0$, and also the energy dependence using $q$, to evaluate the effects that these changes have on both the particle and emission spectra and the size of the PWN.

\subsection{Bulk Particle Motion and Magnetic Field}
The bulk particle speed inside the PWN is parametrised by 
\begin{equation}
V(r) = V_0\left(\frac{r}{r_0}\right)^{\alpha_{\rm{V}}},
\label{V_profile}
\end{equation}
with $\alpha_{\rm{V}}$ the velocity profile parameter. Here $V_0$ is the speed at $r_0$. In modelling the bulk particle motion, the adiabatic energy loss timescale was set constant as done by \cite{Torres2014} as we used their results to calibrate our model. This was done by fixing 
\begin{equation}
\tau_{\rm{ad}} \equiv \frac{E_{\rm{e}}}{\dot{E}_{\rm{ad}}},
\label{eq:ad_parmtr}
\end{equation}
where $\dot{E}_{\rm{ad}} = (\nabla \cdot \mathbf{V})E_{\rm{e}}/3$ and using the analytical form of the term $(\nabla \cdot \mathbf{V})$ that follows from Eq.~\eqref{V_profile}: 
\begin{equation}
(\nabla \cdot \mathbf{V}) = \left( \alpha_{\rm{V}} + 2 \right) \left( \frac{V}{r} \right).
\label{eq_ap:nablaDOTv}
\end{equation}
Thus we find $V_0 = r_0/\tau_{\rm{ad}}$ and $\alpha_{\rm{V}} = 1$ in this case.

We also use a parametrised form of the $B$-field given by 
\begin{equation}
B(r,t) = B_{\rm{age}}\left(\frac{r}{r_0}\right)^{\alpha_{\rm{B}}}\left(\frac{t}{t_{\rm{age}}}\right)^{\beta_{\rm{B}}},
\label{B_Field}
\end{equation}
with $B_{\rm{age}}$ the present-day $B$-field at $r = r_0$ and $t = t_{\rm{age}}$, $t$ the time since the PWN's birth, $t_{\rm{age}}$ is the PWN age, and $\alpha_{\rm{B}}$ and $\beta_{\rm{B}}$ the $B$-field parameters. This parametrised\footnote{We assumed the parametric form for the $B$-field for mathematical expedience assuming the PWN is young. In fact, it is explicitly stated in the \cite{Torres2014} that at earlier times the $B$-field may be approximated using a power law in time ($B \sim t^{-1.3}$; see also \cite{Vorster2013} and references therein. One could use the numerical solution as done by \cite{Torres2014}, but the question then is what the effect of uncertainty on $R_{\rm{PWN}}(t)$ will be on the eventual $B(t)$; i.e., this approach is also not without some assumptions. Our simple approach of using a parametrised $B$-field is meant to be an approximation to the output of a complex MHD code. The solution to such a code is beyond the scope of the current paper and is avoided for the reason that we are focusing on emission physics. In future, one may consider the combination of emission and MHD codes to obtain even more realistic results. To test the effect of parametrising the $B$-field vs.\ calculating it numerically, we implemented the latter approach and found that respective results are very close; see Section~\ref{sec:Cal_Torres}. For older PWNe, a numeric approach will be better and the effect of the reverse shock will also have to be taken into account.} form of the $B$-field goes to infinity if $t=0$ and therefore we limit the $B$-field to $B_{\rm{max}} = 10 B_{\rm{age}}$. Although this is an arbitrary assumption, we found that limiting the $B$-field to larger values ($B_{\rm{max}} = 100 B_{\rm{age}}$ and $B_{\rm{max}} = 1000 B_{\rm{age}}$) has a negligible effect on the predicted SED, but significantly increases the computation time. This parametrised form of the $B$-field is mainly used to see what effect changes in the $B$-field will have on the SED and the size of the PWN. The $B$-field and bulk motion are linked by Faraday's law of induction \citep[e.g.,][]{Ferreira_deJager2008}
\begin{equation}
\frac{\partial \mathbf{B}}{\partial t} = \nabla \times \left(\mathbf{V} \times \mathbf{B}  \right).
\label{eq:B_V}
\end{equation}
The Lorentz force $\mathbf{F} = q(\mathbf{E} + \mathbf{V} \times \mathbf{B})$ is set to zero, assuming that the plasma is a good conductor and thus provides a force-free environment for the leptons. This assumption together with the Maxwell equation 
\begin{equation}
\frac{\partial \mathbf{B}}{\partial t} = -\nabla \times \mathbf{E},
\label{eq:maxwell}
\end{equation}
yields Eq \eqref{eq:B_V}. Assuming that the temporal change of the $B$-field is slow, we set \citep{Kennel1984, Sefako2003, Schock2010}
\begin{equation}
\frac{\partial \mathbf{B}}{\partial t} \simeq 0
\end{equation} 
so that
\begin{equation}
\nabla \times \left(\mathbf{V} \times \mathbf{B}  \right) \simeq 0.
\end{equation}
From this, and assuming spherical symmetry, Eq.~\eqref{eq:B_V} reduces to \citep[e.g.,][]{Kennel1984, Schock2010}
\begin{equation}
VBr={\rm{constant}}=V_0B_0r_0.
\label{eq:vbr=c}
\end{equation}
It can now be shown that by inserting Eq.~\eqref{V_profile} and Eq.~\eqref{B_Field} into Eq.~\eqref{eq:vbr=c}, the following relation holds:
\begin{equation}
\alpha_{\rm{V}}+\alpha_{\rm{B}}=-1.
\label{eq:a_v+a_b=-1}
\end{equation} 
We added the spatial dimension and in this adding two new parameters, $\alpha_{\rm{B}}$ and $\alpha_{\rm{V}}$.  We use the relationship in Eq. \eqref{eq:a_v+a_b=-1} to reduce these two free parameters in our model to one.

\subsection{Numerical Solution to the Transport Equation}
To calculate the numerical solution to the transport equation given in Eq.~(\ref{eq:transportFIN}), we have to discretise this equation. We assume spherical symmetry, thus $\partial/\partial \theta = 0$ and $\partial/\partial \phi = 0$, so that $\nabla^2 N_{\rm{e}} = 1/r^2\left(\partial/\partial r\left[r^2\partial N_{\rm{e}}/\partial r\right]\right)$. We first approached the discretisation process by using a simple Euler method. It soon became clear that this method was numerically unstable. We then decided to use a DuFort-Frankel scheme to discretise Eq.~\eqref{eq:transportFIN} giving
\begin{equation}
\begin{split}
(1-z+\beta)(N_{\rm{e}})_{i,j+1,k} &=2Q_{i,j,1} \triangle t\\
& + (1+z-\beta)(N_{\rm{e}})_{i,j-1,k} \\
& + (\beta+\gamma-\eta)(N_{\rm{e}})_{i,j,k+1} \\
& + (\beta-\gamma+\eta)(N_{\rm{e}})_{i,j,k-1} \\
& -2(\nabla \cdot \mathbf{V})_{i,j,k}\triangle t (N_{\rm{e}})_{i,j,k}\\
& + \frac{2}{\left\{(dE_{\rm{e}})_{i+1,j,k}+(dE_{\rm{e}})_{i,j,k})\right\}} \times\\
& \left\{r_{\rm{a}} \left(dE_{\rm{e,loss}}\right)_{i+1,j,k}(N_{\rm{e}})_{i+1,j,k} \right.\\
& - \left.\frac{1}{r_{\rm{a}}}\left(dE_{\rm{e,loss}}\right)_{i-1,j,k}(N_{\rm{e}})_{i-1,j,k}\right\},
\end{split}
 \label{eq:fin_EQ}
\end{equation}
with $\beta = 2\kappa \Delta t/ (\Delta r)^2$, $\gamma = 2\kappa \Delta t/(r\Delta r)$, $\eta = V_k\Delta t/\Delta r$, $\Delta r$ the bin size of the spatial dimension, $\Delta t$ the bin size of the time dimension, $dE_{\rm{e,loss}}=\dot{E}_{\rm{e,tot}}\Delta t$, and $V_k$ the bulk particle motion in the current radial bin. Also, $r_{\rm{a}} = (\Delta E_{\rm{e}})_{i+1,j,k}/(\Delta E_{\rm{e}})_{i,j,k}$ with
\begin{equation}
z = \left(\frac{1}{(\Delta E_{\rm{e}})_{{i+1,j,k}}-(\Delta E_{\rm{e}})_{{i,j,k}}}\right)\left(\frac{1}{r_{\rm{a}}}-r_{\rm{a}}\right)(\dot{E_{\rm{e}}})_{{i,j,k}}.
\end{equation}
Here $i,j,k$ are the indices for energy, time, and space respectively, with the energy being logarithmically binned, the spatial dimension linearly binned, and the time dimension dynamically binned to optimise the runtime of the code. See \cite{CvR2015} for more details. 

We limit the particle energy following \cite{VdeJager2007},
\begin{equation}
E_{\rm{max}} = \frac{e}{2}\sqrt{\frac{L(t)\sigma}{c(1+\sigma)}}.
\label{eq:E_max_par}
\end{equation}
This is a containment argument, limiting the Larmor radius $r_{\rm{L}} \lesssim 0.5 r_{\rm{s}}$ with $r_{\rm{s}}$ the shock radius. Particles with $E_{\rm{e}}>E_{\rm{max}}$ are assumed to have escaped from the PWN.

\subsection{Boundary Conditions}
The boundary conditions of our model are handled as follows. The multi-zone model divides the PWN into spherical shells to solve Eq.~\eqref{eq:fin_EQ} numerically. The particles are injected into the innermost zone/annulus ($r_{\rm{min}}$) and allowed to propagate through the different zones, with the spectral evolution being governed by Eq.~\eqref{eq:fin_EQ}.  As the initial condition, all zones are assumed to be devoid of any particles, i.e., $N_{\rm{e}} = 0$ at $t = 0$, and a set of ``ghost points" that are also devoid of particles are defined outside the boundaries in time, as the DuFort-Frankel scheme requires two previous time steps. For the spatial dimension, the boundary conditions are reflective at the inner boundary to avoid losing particles towards the pulsar past the termination shock and at the outer boundary $r_{\rm{max}}$ the particles are allowed to escape. To model the escape of particles at the outer boundary, the particle spectrum is set to zero there, while for the reflective inner boundary we need zero flux through the innermost radial shell. Therefore we set
\begin{equation}
 S = -\kappa\left(\frac{(N_{\rm{e}})_{i,j,1} - (N_{\rm{e}})_{i,j,0}}{\Delta r}\right) + V_{i,j,1} \left(\frac{(N_{\rm{e}})_{i,j,1} + (N_{\rm{e}})_{i,j,0}}{2} \right) = 0,
\end{equation}
leading to:
\begin{equation}
(N_{\rm{e}})_{i,j,0} = (N_{\rm{e}})_{i,j,1} \times \frac{\kappa/\Delta r-V_{i,j,1}/2}{\kappa/\Delta r+V_{i,j,1}/2}.
\label{eq:reflective_boundary}
\end{equation}
We solve $N_{\rm{e}}$ for a minimum particle energy of $E_{\rm{min}} = 1 \times 10^{-7}$ erg by allowing particles with smaller energies to escape. The maximum particles energy is limited by Eq.~\eqref{eq:E_max_par}. The injection of particles into the PWN can also be seen as a boundary condition. We inject the particles at a certain rate and assume that the particle injection spectrum $Q^{\prime \prime}$ is uniformly distributed in the first zone. Thus
\begin{equation}
\frac{Q^{\prime \prime}}{V^1_{\rm{shell}}} = Q,
\label{eq:injection_boundary}
\end{equation}
where $V^1_{\rm{shell}}$ is the volume of the first zone and $Q$ the injection spectrum per unit energy, time, and volume as used in Eq.~\eqref{eq:fin_EQ}.

\subsection{Radiation Spectrum}
A time-dependent photon spectrum of each zone can now be calculated, using the electron spectrum $N_{\rm{e}}(r, E_{\rm{e}})$ solved for each zone. For IC we have \citep{Kopp2013}
\begin{equation}
\begin{split}
\left(\frac{dN_\gamma}{dE_\gamma}\right)_{IC} = \,\,\,\,& \frac{g_{IC}}{A} \sum_{l=1}^{3} \int \!\!\!\! \int n_{\varepsilon,l}(r,\varepsilon,T_l) \\
&\times \frac{\mathcal{N}_{\rm{e}}}{\varepsilon E_{\rm{e}}^2}\zeta(E_{\rm{e}},E_\gamma,\varepsilon) d\varepsilon dE_{\rm{e}} , 
\end{split}
\label{eq:ICrad}
\end{equation}
where $A=4\pi d^2$, $d$ the distance to the source, and $\mathcal{N}_{\rm{e}} = N_{\rm{e}}V_{\rm{shell}}$ is the number of electrons per energy in a spherical shell at radius $r$. We consider $l=3$ BB components of target photons, i.e., the cosmic background radiation (CMB), Galactic background infrared (IR) photons, and starlight.

For SR we have
\begin{equation}
\begin{split}
\left(\frac{dN_\gamma}{dE_\gamma}\right)_{\rm{SR}} =\,\,\,\,& \frac{1}{A}\frac{1}{hE_\gamma}\frac{\sqrt{3}e^3B(r,t)}{E_0} \int \!\!\!\! \int_0^{\pi/2} \mathcal{N}_{\rm{e}}(E_{\rm{e}},r)\\
&\times F \left(\frac{\nu}{\nu_{\rm{cr}}(E_{\rm{e}},\alpha,r)}\right)\sin^2 \alpha d\alpha dE_{\rm{e}},
\end{split}
\label{eq:SRrad}
\end{equation}
with $\nu_{\rm{cr}}$ the critical frequency (with pitch angle $\alpha$, which we assume to be $\pi /2$ so that $\sin^2 \alpha = 1$) given by
\begin{equation}
\nu_{\rm{cr}}(E_{\rm{e}},r) = \frac{3ec}{4\pi E_0^3}E_{\rm{e}}^2 B_{\perp}(r,t),
\end{equation}
and
\begin{equation}
F(x) = x\int_x^\infty K_{5/3}(y)dy,
\end{equation}
where $K_{5/3}$ the modified Bessel function of order $5/3$. The total radiation spectrum at Earth is found by calculating Eq.~(\ref{eq:ICrad}) and Eq.~(\ref{eq:SRrad}) for each zone in the model and adding them. 

\subsection{Line-of-Sight Calculation}
Next, the radiation per unit volume can be calculated by dividing the radiation spectrum by the volume of the zone where the radiation originated. This is used to perform the line-of-sight (LOS) calculation to project the radiation onto the plane of the sky in order to find the surface brightness and flux as a function of 2D projected radius. This allows us to estimate the size of the PWN and also study this size as a function of energy. 

\begin{figure}
	\includegraphics[width=\columnwidth]{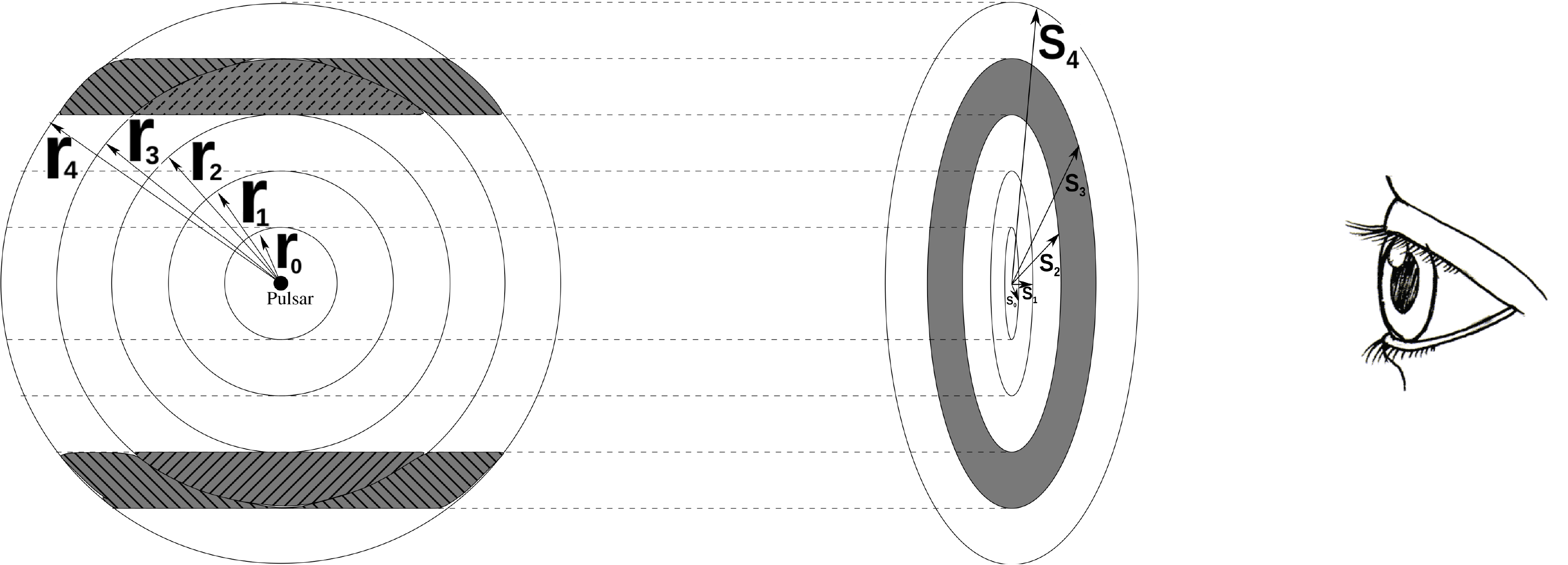}
	\caption{Schematic for the geometry of the LOS calculation.\label{fig:LOS}}
\end{figure}

We multiply the radiation per unit volume by the volume in a particular LOS ($V_{\rm{LOS}}$) as viewed from Earth (Figure~\ref{fig:LOS}). 
The pulsar plus the multi-zone model of the surrounding PWN are on the left hand side of Figure~\ref{fig:LOS} and the right hand side shows how `LOS cylinders' are chosen through the PWN, with the observer looking on from the right. The source is very far from Earth and cylinders instead of cones are chosen as a good first approximation. Cylinders, with radii $s$, intersecting the spherical zones and the spherical shells, with radii $r$, are assumed to have the same bin sizes. This results in the observer viewing the projected PWN as several 2D ``annuli", for example the shaded region in Figure~\ref{fig:LOS}, all with different radii. The radiation in a certain annulus can thus be calculated if the volume of the intersection ($V_{\rm{LOS}}$) between a particular hollow cylinder and the spheres is known. Setting $a \equiv \sqrt{r^2-s^2}$, we find that the volume of intersection between a cylinder and sphere $V(s)$ is
\begin{equation}
\begin{split}
  V(s)& = \frac{4\pi}{3}\left[-\left(r^2-s^2\right)^{\frac{3}{2}} + r^3\right].
\end{split}
\label{eq:intersection}
\end{equation}
The LOS volume ($V_{\rm{LOS}}$) can now be calculated by subtracting the correct volumes from one another. For example, the intersection volume of an annulus with radius $s_k$ and sphere with $r_i$ is  
\begin{equation}
  V_{\rm{LOS}} = \left(V_{i,k} - V_{i,k-1}\right) - \left(V_{i-1,k} - V_{i-1,k-1}\right).
\label{eq:washer}
\end{equation}
This expression, however, holds only when $s < r$. If $s$ is larger or equal to $r$, then the intersection volume will simply be the volume of the sphere of radius $r$. The total radiation for the specific LOS, or annulus, can be calculated by adding the radiation for all the segments (Figure \ref{fig:LOS}). To find the total radiation at Earth from the PWN, the radiation from all the different LOSs (annuli) may be added. 

As a test of this LOS calculation, we summed the total flux from all the spheres to find the total flux from the PWN and then also added the flux from all the cylinders after the LOS calculation. Both these calculations yielded the same flux. We can now use this projected flux to calculate the surface brightness profile and thus calculate the size of the PWN (Section \ref{sec:Space}).

\section{Code Calibration via SED fits}
\label{sec:cal}
PWN G0.9+0.1 will be used as a case study to calibrate our newly developed code and here we briefly summarise some of its observational properties. \cite{HelfandB1987} observed G0.9+0.1 for 45-minute integrations at 20~cm and 6~cm, which led to the discovery of the composite nature of this bright, extended source near the Galactic Centre (GC) in the radio band. SNR G0.9+0.1 has since become a well-known supernova remnant, with an estimated age of a few thousand years. This source exhibits a flat-spectrum radio core ($\sim2'$ across) corresponding to the PWN, and also clearly shows steeper shell components ($\sim 8'$ diameter shell). While performing a survey on the GC, \cite{Sidoli2004} serendipitously observed SNR G0.9+0.1 using the \textit{XMM-Newton} telescope. Their observations provided the first evidence of X-ray emission from PWN G0.9+0.1. \cite{Sidoli2004} fit an absorbed power-law spectrum that yielded a photon index of $\Gamma \sim 1.9$ and an energy flux of $F = 4.8 \times 10^{-12}$ erg cm$^{-2}$ s$^{-1}$ in the 2$-$10 keV energy band. This translates to a luminosity of $L_{\rm X} \sim 5 \times 10^{34}$ erg s$^{-1}$ for a distance of 10 kpc. \cite{G0.9+0.1_HESS} studied VHE $\gamma$ rays from the GC with the H.E.S.S.\ telescope. During the observation of Sgr A$^*$, two sources of VHE gamma rays were clearly visible, with SNR G0.9+0.1 being one of these sources. They performed a power-law fit to the observed spectrum and found a photon index of $2.29 \pm 0.14_{\rm{stat}}$ with a photon flux of $(5.5 \pm 0.8_{\rm{stat}})\times 10^{-12}$~cm$^{-2}$~s$^{-1}$ for energies above 200 GeV. This flux is only $\sim 2 \%$ of the flux from the Crab Nebula, making PWN~G0.9+0.1 one of the weakest sources detected at TeV energies to date. Some years later, the radio pulsar PSR~J1747$-$2809 was discovered in PWN~G0.9+0.1 with period\footnote{Below we discuss calibration of our model against that of \cite{VdeJager2007}. To closer align with their procedure, for the sake of calibration, we fixed the value of $L_0$ and birth period $P_0 = 43$ ms \citealt{Swaluw2001}, assuming no decay of the pulsar $B$-field, i.e., $P_0 \dot{P}_0=PP_0$. In the rest of the paper, however, we calculate $\tau_{\rm{c}}$ using $P$ and $\dot{P}$, we assume $t_{\rm{age}}$, and from this follows $\tau_0$ and $L_0$ (without the need to calculate $P_0$ and $\dot{P}_0$ explicitly).} $P = 52$~ms and $\dot{P} = 1.85\times10^{-13}$ \citep{G0.9+0.1data}.

In the next section, we calibrate our new model against a previous more basic model \citep{VdeJager2007}. This model only assumed a parametric form for the $B$-field, and did not take into account work done by the $B$-field and the effect thereof on its time dependence. The calibration with this older model is a first point of reference and is also done for historical reasons, since our new model incorporates many of the basic elements of the \cite{VdeJager2007} model. We also calibrated our new model against a more modern model \citep{Torres2014}. Both of these earlier works assumed one-zone models (no spatial dependence). We decided to add another calibration using the model of \citet[][results not shown since we focused on PWN G0.9+0.1]{Lu2017}. The fact that the respective predicted spectra are in reasonable agreement increases our confidence in the accuracy of our model.

\subsection{Calibration against the Model of \citet{VdeJager2007}}
\begin{figure}
\includegraphics[width=\columnwidth]{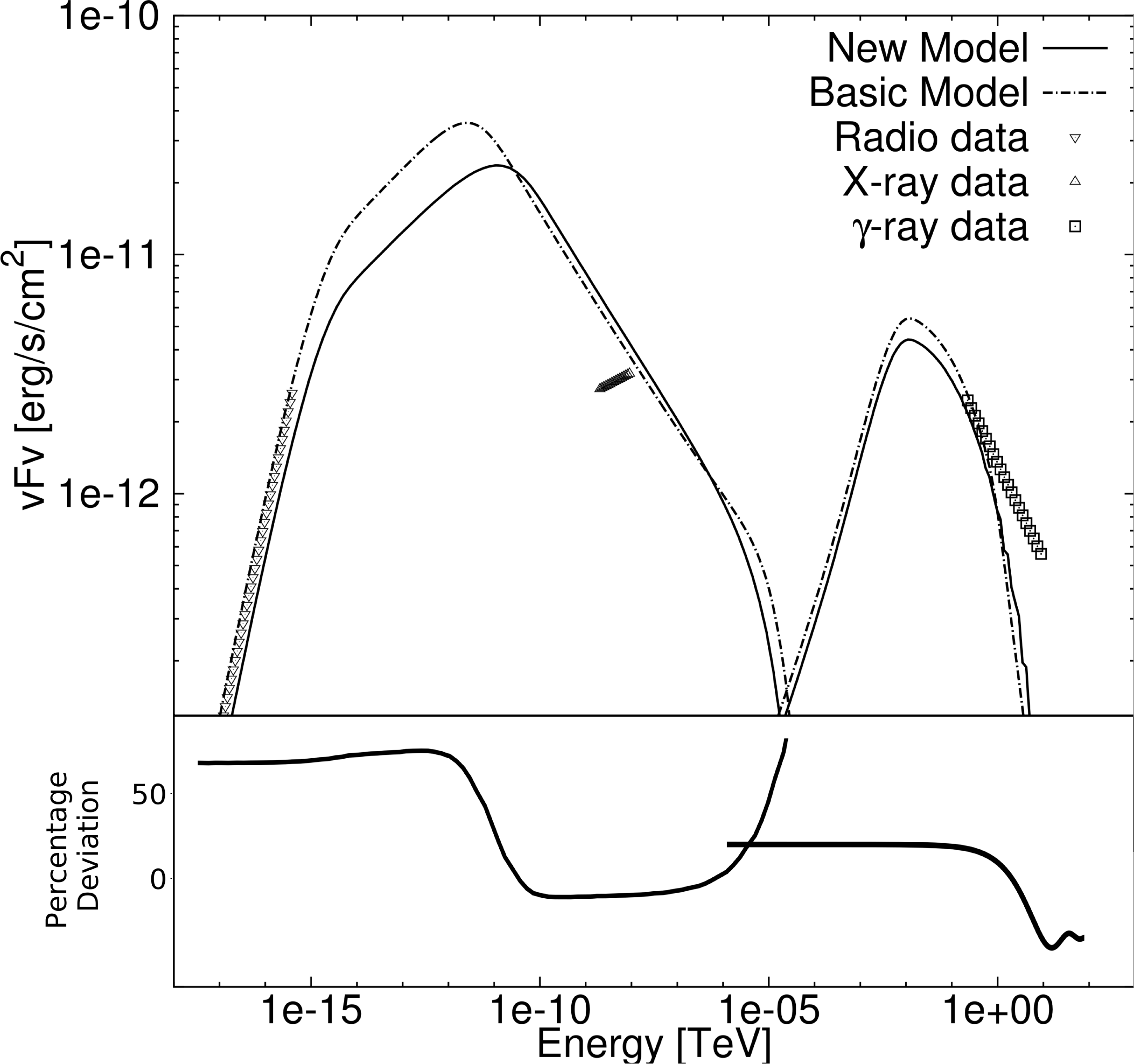}
\caption{Calibration against the model of \citet{VdeJager2007} for PWN G0.9+0.1\label{fig:Calibrate}. Bottom panel indicates the percentage deviation between the two SEDs.}
\end{figure}

The assumed model parameters used to calibrate our model against that of \cite{VdeJager2007} are listed in Table~\ref{tbl:G0.9}. In Table~\ref{tbl:G0.9}, $n$ is the braking index given by $n = \Omega \ddot{\Omega}/\dot{\Omega}^2$, with $\Omega = 2 \pi /P$ the angular speed and $P$ the period of rotation of the pulsar; $\beta_{\rm{VdJ}}$ is the $B$-field parameter as in Eq.~\eqref{eq:fieldCV}, and $B(t_{\rm{age}})$ is the present-day $B$-field. In this first calibration with \cite{VdeJager2007}, we use $B(t_{\rm{age}})=40.0~\mu$G, noting that their model was developed before the discovery of PSR J1747$-$2809 associated with PWN G0.9+0.1. The more reasonable value for the present-day $B$-field, 14.0 $\mu$G, is used in the calibration against the model of \cite{Torres2014} in the next section as we now know $P$ and $\dot{P}$ for the embedded pulsar, as mentioned above. Also, $\epsilon$ is the conversion efficiency as mentioned in Eq.~\eqref{normQ}, $t_{\rm{age}}$ is the age of the PWN, $\tau_0$ is the characteristic spin-down timescale of the pulsar, $d$ is the distance to the PWN, $\alpha_1$ and $\alpha_2$ are the spectral indices, and $L_0$ the birth spin-down luminosity. The sigma parameter ($\sigma$) is the ratio of the electromagnetic to particle energy density and is used to calculate the maximum particle energy. We chose three soft-photon components: the CMB with a temperature of $T_1 = $ 2.76~K and an average energy density of $u_1 = $ 0.23~eV/cm$^3$, Galactic background infrared photons as component 2, and optical starlight as component~3 (with $T_i$ and $u_i$ as given in Table~\ref{tbl:G0.9}). For these assumed model parameters we find the SED as shown in Figure~\ref{fig:Calibrate}. The radio data are from \cite{HelfandB1987}, the $X$-ray data from \cite{Porquet2003} and \cite{Sidoli2004}, and the gamma-ray data from \cite{G0.9+0.1_HESS}. The solid line represents our predicted SED while the dashed line shows the output from the model of \cite{VdeJager2007}.

To compare our new model to the model of \cite{VdeJager2007}, we had to remove the effects of the bulk particle motion, as their model did not incorporate such motion and only considered diffusion, SR losses, and particle escape. Thus their model did not include adiabatic losses nor convection (see below). The way the effects of these processes are removed from the new model is by simply setting the bulk speed inside the PWN to zero. \cite{VdeJager2007} also modelled the $B$-field by parametrising it as
\begin{equation}
B(t) = \frac{B_0}{1+\left( t / \tau_0 \right)^{\beta_{\rm{VdJ}}}}.
\label{eq:fieldCV}
\end{equation}
Our model was adapted to also parametrise the $B$-field using this same time-dependent form. These two simple changes to our model allowed us to calibrate our model against theirs as seen in Figure~\ref{fig:Calibrate}. In Table \ref{tbl:G0.9} the present-day $B$-field $B_{\rm{age}}$.

\begin{table}
\begin{center}
\caption{Values of model parameters as used in the calibration against the model of \citet{VdeJager2007} for PWN G0.9+0.1.\label{tbl:G0.9}}
\resizebox{\columnwidth}{!}{
\begin{tabular}{crr}
\hline
Model Parameter & Symbol & Value \\
\hline
Braking index & $n$ & 3  \\
$B$-field parameter & $\beta_{\rm{VdJ}}$ & 0.5 \\
Present-day $B$-field & $B(t_{\rm{age}})$ & 40.0 $\mu \rm{G}$ \\
Conversion efficiency  &    $\epsilon$    &  0.6\\
Age & $t_{\rm{age}}$ & 1~900 yr\\
Characteristic timescale & $\tau_0$ & 3~681 yr\\
Distance & $d$ & 8.5 kpc  \\
$Q$ index 1& $\alpha_1$& -1.0 \\
$Q$ index 2& $\alpha_2$& -2.6 \\
Initial spin-down power($10^{38}\rm{erg}$ $\rm{s^{-1}}$)& $L_0$ & 0.99  \\ 
Sigma parameter & $\sigma$ & 0.2 \\
Soft-photon component 1 & $T_1$ and $u_1$ & $2.76$ K, $0.23~\rm{eV/cm^3}$ \\
Soft-photon component 2 & $T_2$ and $u_2$ & $35$ K, $0.5~\rm{eV/cm^3}$ \\  
Soft-photon component 3 & $T_3$ and $u_3$ & $4~500$ K, $50~\rm{eV/cm^3}$ \\
\hline
\end{tabular}
}
\end{center}
\end{table}

Our time-dependent, multi-zone PWN model does not reproduce the results of \cite{VdeJager2007} exactly, but the SEDs are quite close. The reason for this is the fact that the older model did not take into account IC losses in the particle transport, since it assumed SR losses to dominate. This led to particle energy losses being underestimated, leaving an excess of high-energy particles. Their IC radiation is therefore slightly higher than our new model prediction. Other differences may result from our very different treatment of the particle transport as we solved a full transport equation and \cite{VdeJager2007} solved a linearised transport equation using energy losses, diffusion and effective timescales. 

One thing to note here is that in Table~\ref{tbl:G0.9}, the two variables $\epsilon$ and $\sigma$ are independent. They are, however, in reality related by $\epsilon = 1/(1+\sigma)$. This inconsistency is only present in the calibration with \cite{VdeJager2007} and is correctly implemented in the rest of the paper.

Our model fits the data quite well, but still has trouble in fitting the slope of the X-ray spectrum. \cite{Vorster2013} modelled PWN G21.5$-$0.9 where they also encountered this problem when using a broken-power-law injection spectrum that connects smoothly at some break energy. They therefore used a two-component particle injection spectrum that does not transition smoothly (instead the low-energy component cuts off steeply in order to connect to the lower-flux, high-energy component), allowing them to fit both the radio and X-ray spectral slopes. This is something worth noting for future development of our code. 

\subsection{Calibration against the Model of \citet{Torres2014}}\label{sec:Cal_Torres}
\begin{figure}
\includegraphics[width=\columnwidth,keepaspectratio]{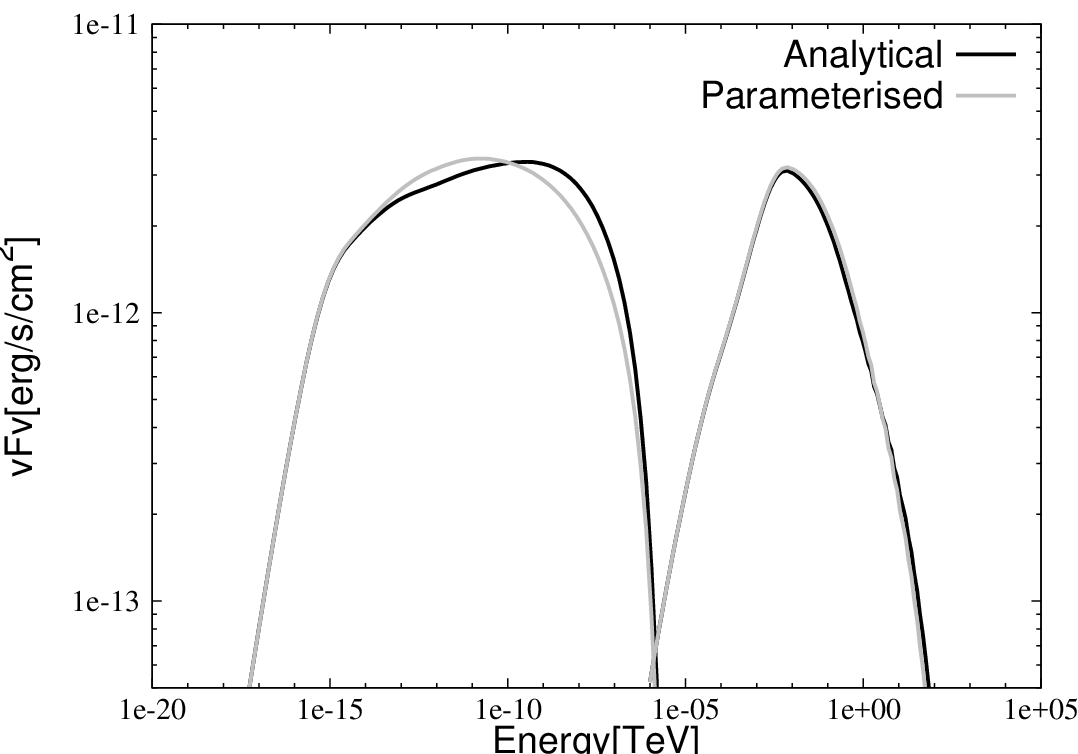}
\caption{Comparison of the predicted SED for the parametric vs.\ analytical treatment of the temporal evolution of the $B$-field.\label{fig:magneticTorres}}
\end{figure}

As a second calibration, we used results from a more recent study by \cite{Torres2014}, who created a time-dependent model of young PWNe by modelling them as a single sphere. We again use PWN G0.9+0.1 as the calibration source. The assumed model parameters for this second calibration are given in Table~\ref{tbl:G0.9_Torres}. The $B$-field is now modelled according to Eq.~\eqref{B_Field}, hence the values of $\alpha_{\rm{B}}$ and $\beta_{\rm{B}}$ in Table~\ref{tbl:G0.9_Torres}. Some of the parameters are different from those used during the calibration with the model of \cite{VdeJager2007}. One of these changes is the present-day $B$-field that is now set to $14~\mu G$, versus the previous value of $40~\mu G$. Furthermore, the discovery of pulsar PSR J1747$-$2809 in the PWN G0.9+0.1 yielded $P$ and $\dot{P}$ which pin down the value of $L(t_{\rm{age}})$; see the Appendix. The $B$-field is parametrised using $\alpha_{\rm{B}} = 0$ and $\beta_{\rm{B}} = -1.3$, which, from Eq.~\eqref{B_Field}, indicates that the $B$-field is constant in the spatial dimension. \cite{Torres2014} model the time dependence of the $B$-field using
\begin{equation}
\int_0^t (1-\epsilon) L(t')R_{\rm{PWN}}(t')dt' = W_{B}R_{\rm{PWN}},
\label{eq:Torres_B}
\end{equation}
where
\begin{equation}
W_{B}=\frac{4\pi}{3}R^3_{\rm{PWN}}(t)\frac{B^2(t)}{8\pi},
\label{eq:Torres_B2}
\end{equation}
and mention that if the age of the PWN is less than the characteristic age ($t_{\rm{age}} < \tau_0$), then $B(t) \propto t^{-1.3}$. Therefore we set the value of $\beta_{\rm{B}} = -1.3$. While \cite{Torres2014} solved $B(t)$ numerically, we can approximate the early-age limit of $B(t)$ using such a power law. One thing to note here is the usage of $R_{\rm{PWN}}$. \cite{Torres2014} explicitly use a time-dependent PWN radius for G0.9+0.1, setting $R_{\rm{PWN}}(t_{\rm{age}}) = 2.5$ pc. However, we do not. Instead we choose\footnote{For simplicity we assume this distant boundary for the PWN where particles escape. We restrict ourselves to modelling young PWN where free expansion of the wind should be justified. Later evolutionary phases may be characterised by a reverse shock, or reverberation phase where interaction with the ambient medium is much more important. If we would enforce particle escape at a moving boundary $R_{\rm{PWN}}$ rather than following our approach of escape at a distant boundary, the particle density may be somewhat lower, leading to slightly lower fluxes than predicted by our model.} an escape boundary $r_{\rm{max}} \gg R_{\rm{PWN}}$, and then later calculate the observable size of the PWN by noting where the surface brightness has decreased by two thirds from the value at $r_{\rm{min}}$ (from an observer's point of view; this is possible since we have information about the morphology of the PWN). Our approach is admittedly different from the standard one, but with a very particular motivation: If we assume a standard expression for $R_{\rm{PWN}}(t)$ and if the age of the PWN is much smaller than the radiation loss timescale, one would expect no evolution of PWN size with energy, contrary to what is observed in some PWNe\footnote{ Whether a PWN's morphology is energy-dependent seems to be closely linked with the evolutionary stage of the PWN and to whether particles efficiently escape from the system beyond $R_{\rm PWN}(t)$. As mentioned in the Introduction (Section~\ref{sec:intro}), young systems with slow moving embedded pulsars may manifest as composite SNRs with a high degree of spherical symmetry, prior to the interaction with the SNR reverse shock. In such systems, particles may not have had time to cool significantly due to radiation losses, leading to a morphology that seems to be largely energy-independent. Middle-aged PWNe may exhibit complex morphologies (e.g., collimated X-ray emission vs. more diffuse ambient radio emission), while in older PWNe, the $\gamma-$ray emission may dominate the radio and X-ray emission. This is probably due to the fact that high-energy particles are the ones that preferentially cool and escape from the PWN. \citet{Hinton2011} argue that while confinement of particles in PWNe may be effective during the early stages of evolution, the interaction with the SNR reverse shock may disrupt the PWN via, e.g., the growth of Rayleigh-Taylor instabilities, and diffusion of particles out of the PWN becomes possible. In the case of PWN G0.9+0.1, we may be witnessing an intermediate case. While this PWN is quite young and the radio and X-ray sizes are very similar, the X-ray morphology seems to be slightly smaller than the radio (\citealt{Dubner2008}). This hint of morphological evolution (Figure \ref{fig:SvE_combine}) is consistent with the observed softening of the X-ray spectrum as one moves from the inner to outer regions of the PWN, indicating the effect of SR losses in this system; \citealt{Porquet2003}).}. Conversely, we calculate the energy-dependent PWN size using the predicted surface brightness profile. However, this approach does 
not ignore the dynamical evolution of the PWN. While we determine $R_{\rm{PWN}}(t)$ from the emission properties, we do take into account the effect of evolution on the $B$-field profile by choosing $\beta_{\rm{B}} = -1.3$.  Our parametric approach captures the essence of the evolution (e.g., as assumed by \citealt{Torres2014}) in a simplified way, but allows us the freedom to infer this profile, should the data require a somewhat different behaviour for the decline of the B-field with radius. As a test we performed alternative runs of our code, in which we included the formalism of \citet{Torres2014} to calculate the $B$-field. We found no significant difference in the predicted SED when using these two different approaches (Figure~\ref{fig:magneticTorres}), justifying our usage of the parametric approach when modelling young PWNe.

\begin{table}
\begin{center}
\caption{Values of model parameters as used in the calibration against the model of \citet{Torres2014} for PWN G0.9+0.1.\label{tbl:G0.9_Torres}}
\resizebox{\columnwidth}{!}{
\begin{tabular}{crr}
\hline
Model Parameter & Symbol & Value \\
\hline
Braking index & $n$ & 3  \\
$B$-field parameter & $\alpha_{\rm{B}}$ & 0.0 \\
$B$-field parameter & $\beta_{\rm{B}}$ & -1.3 \\
$V$ parameter & $\alpha_{\rm{V}}$ & 1.0 \\
Present-day $B$-field & $B(t_{\rm{age}})$ & 14.0 $\mu \rm{G}$ \\
Conversion efficiency  &    $\epsilon$    &  0.99\\
Age & $t_{\rm{age}}$ & 2~000 yr\\
Characteristic timescale & $\tau_0$ & 3~305 yr\\
Distance & $d$ & 8.5 kpc  \\
$Q$ index 1& $\alpha_1$& -1.4 \\
$Q$ index 2& $\alpha_2$& -2.7 \\
Initial spin-down power($10^{38}\rm{erg}$ $\rm{s^{-1}}$)& $L_0$ & 1.1  \\ 
Sigma parameter & $\sigma$ & 0.01 \\
Soft-photon component 1 & $T_1$ and $u_1$ & $2.76$ K, $0.23~\rm{eV/cm^3}$ \\
Soft-photon component 2 & $T_2$ and $u_2$ & $30$ K, $2.5~\rm{eV/cm^3}$ \\  
Soft-photon component 3 & $T_3$ and $u_3$ & $3~000$ K, $25~\rm{eV/cm^3}$ \\
\hline
\end{tabular}
}
\end{center}
\end{table}

\begin{figure}
\includegraphics[width=\columnwidth,keepaspectratio]{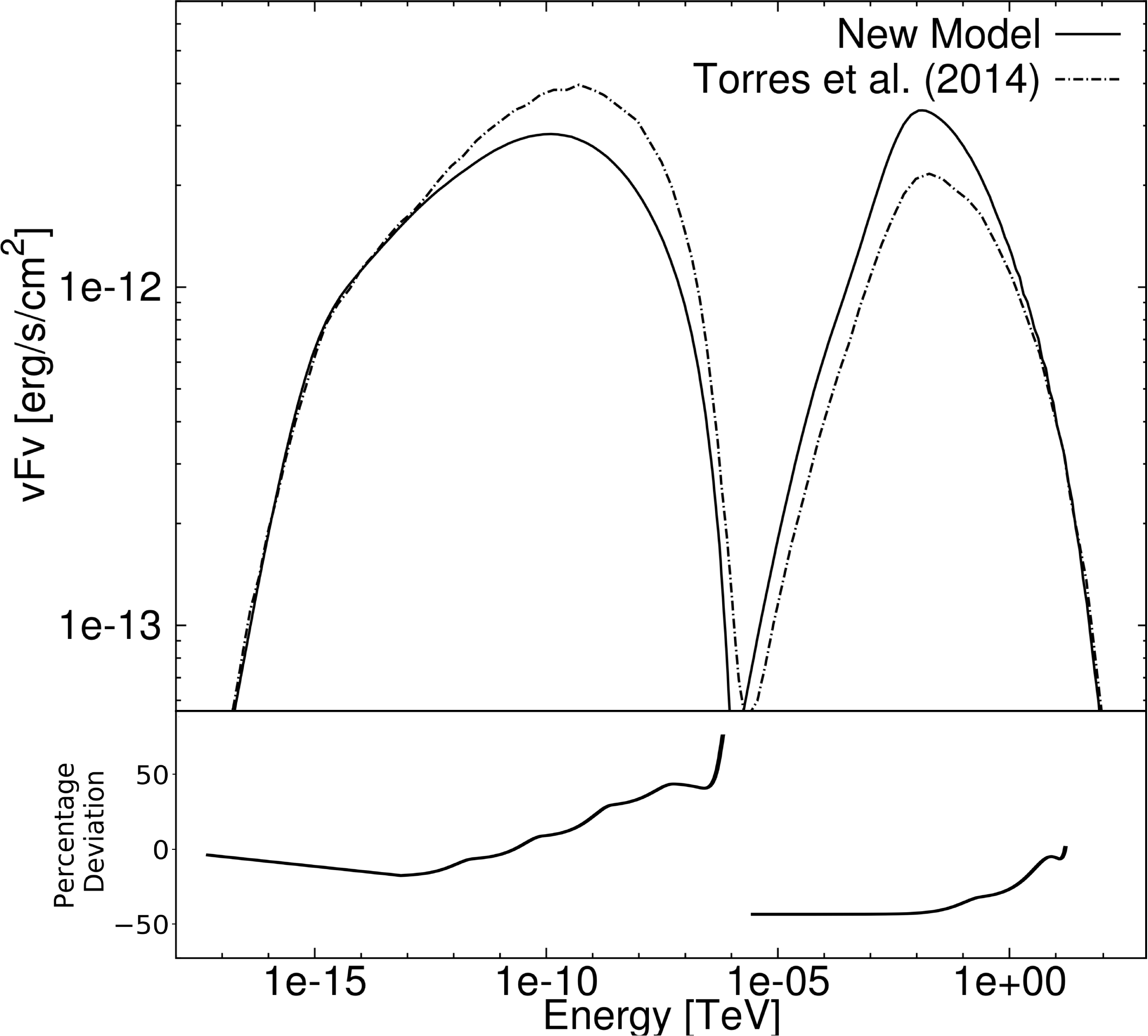}
\caption{Calibration against the model of \citet{Torres2014} for PWN G0.9+0.1. Bottom panel indicates the percentage deviation between the two SEDs.\label{fig:Calibrate_Torres}}
\end{figure}

The bulk motion of the particles is parametrised by Eq.~\eqref{V_profile} using model parameters $\alpha_{\rm{V}}$, $V_0$, and $r_0 = r_{\rm{min}}$ and the velocity is parametrised by setting $\alpha_{\rm{V}} = 1.0$, with $V_0=r_0/t_{\rm{age}}$. This is done so that our model can have the same adiabatic energy loss rate assumed by \cite{Torres2014}. They have a constant adiabatic energy loss timescale and to reproduce this in our model, we have to set $\alpha_{\rm{V}} = 1$ (see Eq. [\ref{eq_ap:nablaDOTv}]). This leads to a value for $V_0$ from the adiabatic timescale:
\begin{equation}
\tau_{\rm{ad}} = \frac{E}{\dot{E}_{\rm{ad}}},
\label{eq:ad_parmtr2}
\end{equation}
where $\dot{E}_{\rm{ad}} = (\nabla \cdot \mathbf{V})E_{\rm{e}}/3$. By using the analytical form of $(\nabla \cdot \mathbf{V})$ in Eq.~\eqref{eq_ap:nablaDOTv} we find that $V_0 = r_0/\tau_{\rm{ad}}$. This is, however, not physical, if the relationship between $V(r)$ and $B(r,t)$ in Eq.~\eqref{eq:a_v+a_b=-1} holds. From these equations it is clear that $\alpha_{\rm{V}} = -1$ when $\alpha_{\rm{B}} = 0$. The conversion efficiency ($\epsilon$) is very large, but there exists a degeneracy between $\epsilon$ and $L_0$ and therefore this is still a preliminary value. The changes in $B(r,t)$ and $V(r)$ are the only substantial differences between the model of \cite{Torres2014} and our model. The rest of the parameters are very similar to the previous case, e.g., the indices of the injection spectrum and the soft-photon components used in the calculation of the IC spectrum.

Figure~\ref{fig:Calibrate_Torres} compares our predicted SED with the model prediction of \cite{Torres2014}, with their results shown by the dashed-dotted line and our model SED shown as the solid line. The differences between the two models stem from the different ways in which the transport of particles is handled. In our code we incorporated a Fokker-Planck-type transport equation and \cite{Torres2014} modelled the transport by using average timescales.

During the calibration of the code, other sources where also modelled (e.g., G21.5-0.9, G54.1+0.3, and HESS J1356-645). We found that the model yields reasonable fits for most of the chosen sources as long as they are young PWNe. These results will be shown in a subsequent paper  where we will perform a more detailed PWN population study.

\section{Parameter Study}
\label{sec:Par}
We can now investigate the effects of several of the free model parameters on the predicted particle spectrum and SED. As a reference model for this section, we use the same parameters that were used in the calibration against \cite{Torres2014} for G0.9+0.1, as in Figure~\ref{fig:Calibrate_Torres}. The SED of the PWN is calculated at Earth for each spherical zone and then these are added to find the total flux from the PWN. 

\begin{figure}
\includegraphics[width=\columnwidth]{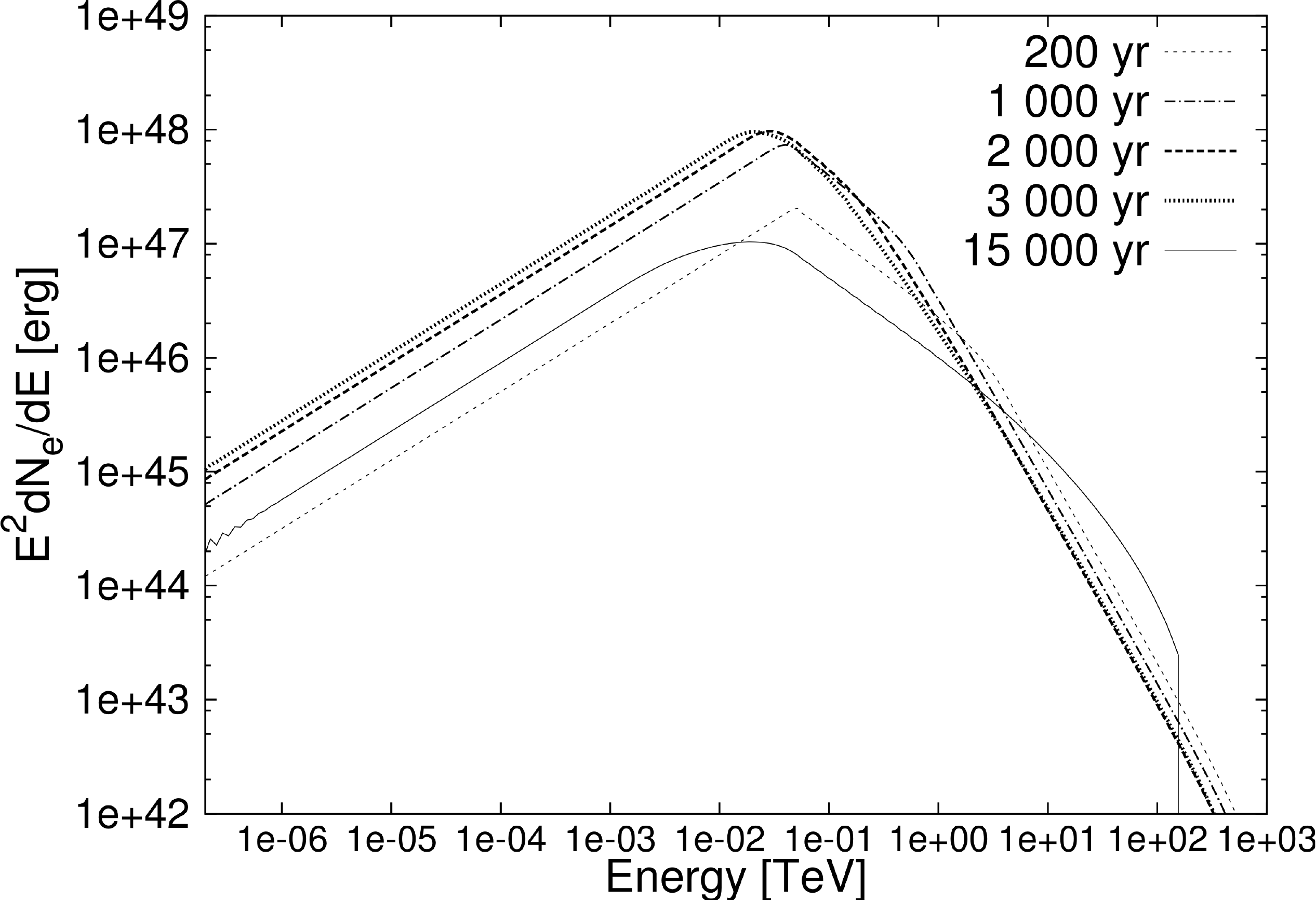}
\caption{\label{fig:dnde_time}Time evolution of the lepton spectrum.}
\end{figure}

\subsection{Time Evolution (age)}
In Figure~\ref{fig:dnde_time} the time evolution of the lepton spectrum is shown. From this Figure it can be seen that when the PWN is still very young ($t_{\rm{age}}\sim 200$ yr) the particle spectrum closely resembles the shape of the injection spectrum apart from the spectral break at a few TeV that develops due to radiation losses. As the PWN ages, however, it starts to fill up with particles (giving an increased $E^2_{\rm{e}}dN_{\rm{e}}/dE_{\rm{e}}$) and at some stage the PWN is totally filled, at an age in the order of a few thousand years. After this the level of the particle spectrum decreases. This is due to the particles losing energy over time due to SR, IC, adiabatic energy losses, and escape, and also due to the fact that the embedded pulsar is spinning down, resulting in fewer particles being injected into the PWN. The effect of the spun-down pulsar can be clearly seen in Figure~\ref{fig:dnde_time} by observing the spectrum at 15~000~yr. By this time the embedded pulsar has significantly spun down so that the total particle spectrum is lower than it was at $\approx 200$~yr due to the fact that now more particles are escaping from the modelled region at $r_{\rm{max}}$ than are being injected by the pulsar. Also note the leftward shift of $E_{\rm{b}}$ due to radiative losses. The bump at high energies for 15~000~yr is due to a pile-up of particles. This occurs due to the decreased $B$-field $B(t)$, resulting in an increased diffusion coefficient and also decreased SR energy losses. These losses are energy-dependent and therefore the high-energy particles will be most affected. The increased diffusion will cause the particles to resemble the injection spectrum more and more due to suppressed SR losses. 

The particle spectrum in Figure~\ref{fig:dnde_time} not only goes up and down as the PWN ages, but the whole spectrum shifts to lower energies. This can be seen by looking at where the spectrum peaks and also at the tails at high and low energies. This is due to the fact that the particles lose energy through previously mentioned mechanisms. Due to the SR energy losses, the particle spectrum will develop a high energy break at some break energy. By using $\dot{E}_{\rm{SR}}$ as in Eq. \eqref{SR}, we can calculate the timescale for synchrotron losses ($\tau_{\rm{SR}}$) and by setting it equal to the age of the PWN ($t_{\rm{age}}$), one may estimate where this second break is expected in the spectrum:
\begin{equation}
  \tau_{\rm{SR}} = \frac{E_{\rm{e}}}{\dot{E}_{\rm{SR}}} = t_{\rm{age}} \Rightarrow E_{\rm{e}} \propto \frac{1}{t_{\rm{age}} \langle B \rangle^2}.
  \label{eq:SR_timescale}
\end{equation}
Thus from Eq.~\eqref{eq:SR_timescale} we can see that the break should move to lower energies as the PWN ages. In Eq.~\eqref{eq:SR_timescale} we have to use the average $B$-field $\langle B \rangle$ over the lifetime of the PWN as the present-day $B$-field is too small. This is visible in Figure~\ref{fig:dnde_time} where the break for 200~yr is at $\approx$~2~TeV, for 1~000~yr at~$\approx$~0.6~TeV, for 2~000~yr at~$\approx$~0.2~TeV, and for 5~000~yr at~$\approx$~0.15~TeV. By inserting these values into Eq.~\eqref{eq:SR_timescale} we find a reasonable value of $\langle B \rangle \sim 150~\mu$G. These results are similar to those found by \cite{Torres2014}.

\subsection{Magnetic Field}
The $B$-field $B(r,t)$ inside the PWN (which determines the diffusion) plays a large role in determining the shape of the SED (level and break energy of SR and IC component), and is characterised by the free parameters $B_{\rm{age}}$, $\alpha_{\rm{B}}$ and $\beta_{\rm{B}}$ (Table~\ref{tbl:G0.9_Torres}). As a default, the present-day $B$-field is set to $14~\mu G$ and is then changed to $10~\mu G$, $20~\mu G$, and to $40~\mu G$ to see what effect this will have. 
\begin{figure}
\includegraphics[width=\columnwidth]{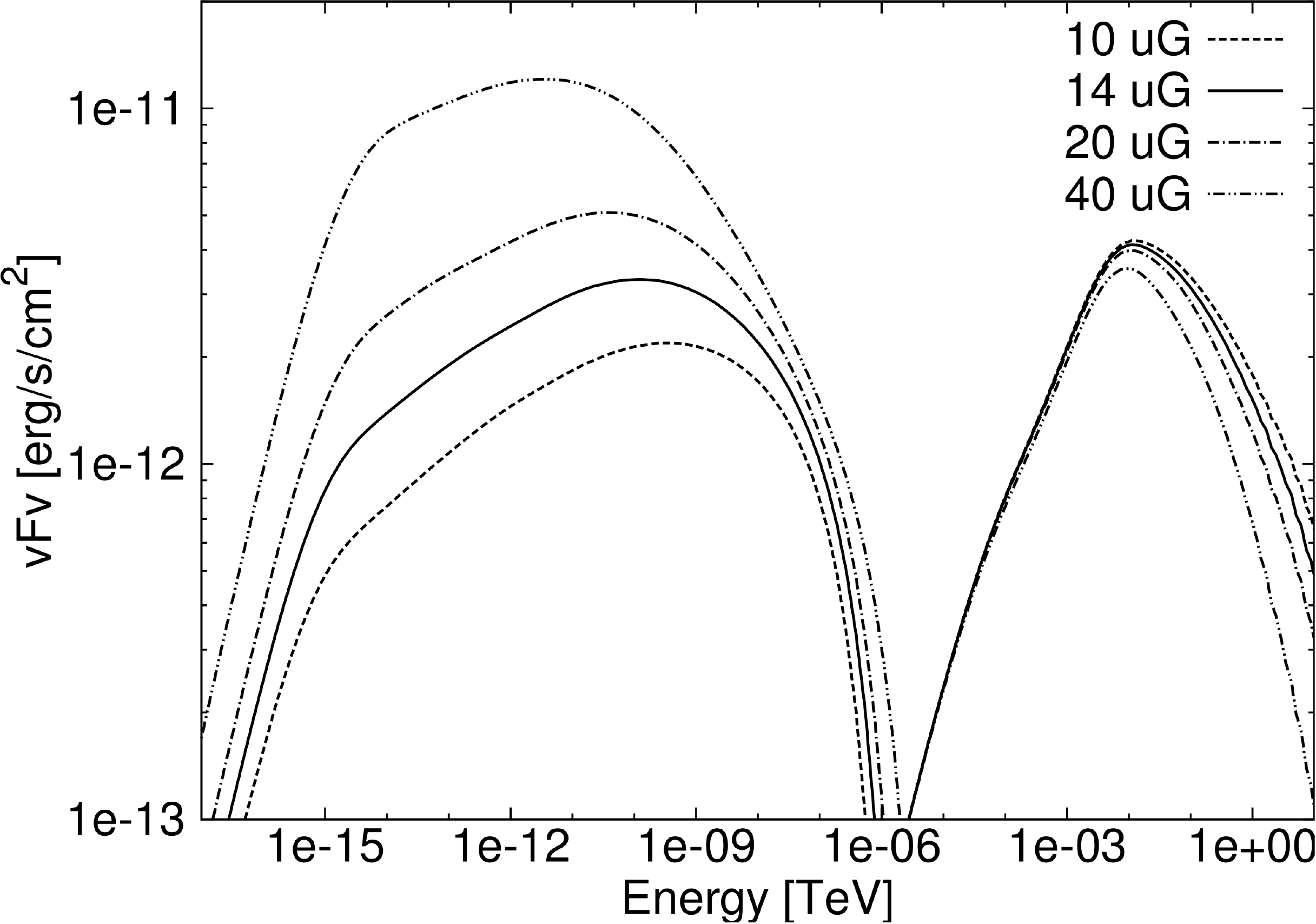}
\caption{\label{fig:SED_Change_B}SED for PWN~G0.9+0.1 with a change in the present-day $B$-field.}
\end{figure}
Here, we fix the values for $\alpha_{\rm{B}}$ and $\beta_{\rm{B}}$ to 0.0 and -1.3, respectively, as mentioned earlier, so only the value of $B_{\rm{age}}$ was changed (see Section~\ref{sec:alph_Valph_B} for a discussion on the changes in $\alpha_{\rm{V}}$ and $\alpha_{\rm{B}}$). As the $B$-field increases from 10~$\mu G$ to 40~$\mu G$, the particle spectrum becomes softer at high energies, since $\dot{E}_{\rm{SR}} \propto E_{\rm{e}}^2B^2$. Thus higher-energy particles lose more energy so that there are fewer particles at high energies left to radiate. The high-energy tail of the IC spectrum in Figure~\ref{fig:SED_Change_B} is therefore lower for a larger $B$-field. The SR power is directly proportional to the $B$-field strength squared and thus as the $B$-field increases, so does the SR.

\subsection{Bulk Particle Motion}
\begin{figure}
\includegraphics[width=\columnwidth]{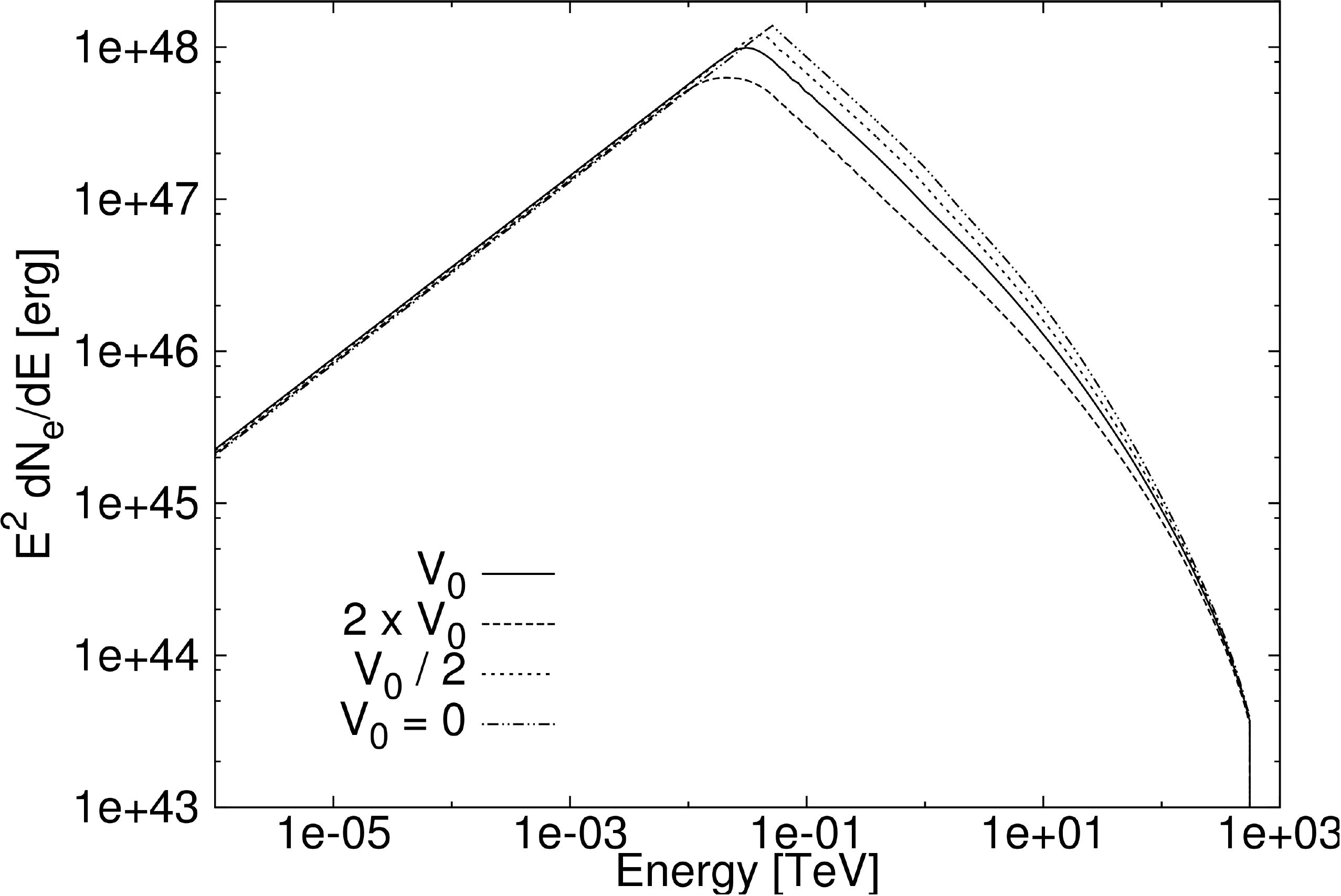}
\caption{\label{fig:Par_Change_V}Particle spectrum for PWN~G0.9+0.1 for a change in the bulk speed of the particles.}
\end{figure}

\begin{figure}
\includegraphics[width=\columnwidth]{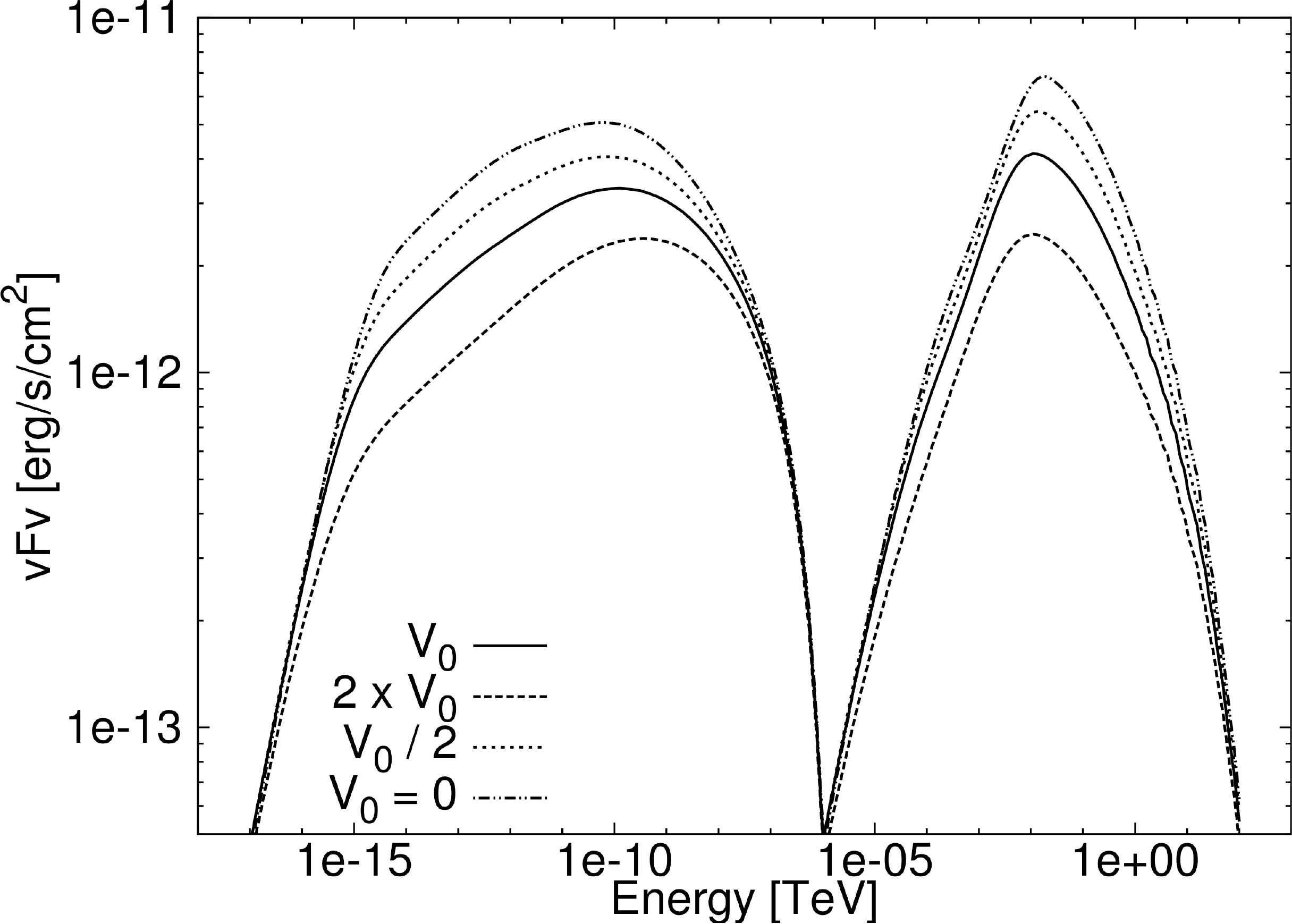}
\caption{\label{fig:SED_Change_V}SED for PWN~G0.9+0.1 for a change in the bulk speed of the particles.}
\end{figure}

The bulk particle motion (particle speed) in the PWN is modelled by Eq. (\ref{V_profile}) and the value for $\alpha_{\rm{V}} = 1$ is kept constant here, although the value of $V_0$ is changed to $V_0 = 0$, $2V_0$ and $V_0/2$ as can be seen in Figure \ref{fig:Par_Change_V} and \ref{fig:SED_Change_V}. To compare our results with those of \cite{Torres2014} we need the same form for the bulk particle motion (see Eq. [\ref{eq:ad_parmtr2}]). The adiabatic timescale that \cite{Torres2014} used for G0.9+0.1 was $\sim$~2~000~yr, giving $V_0 = 5 \times 10^{-5}$~pc/yr for $r_0 = 0.1$~pc and $\tau_{\rm{ad}} = 2~000$~yr.

In Figure~\ref{fig:Par_Change_V} the particle spectrum increases as $V_0$ is lowered. This is due to the fact that for a lower speed, the particles lose less energy due to adiabatic losses. The adiabatic energy losses also account for the leftward shift of the peak in the particle spectrum. The radiation spectrum is linked to the particle spectrum and therefore a lower particle spectrum results in a lower radiation spectrum. This effect can be seen in Figure~\ref{fig:SED_Change_V} where the radiation decreases with an increase in the bulk speed of the particles. For high energies, SR energy losses dominate over adiabatic losses, and therefore the high-energy tail of the radiation spectrum is independent of changes in $V_0$ and the tails converge.

\subsection{Injection Rate / Initial Spin-Down Rate}
The particles in the PWN are injected from the embedded pulsar and the injected spectrum is normalised using the time-dependent spin-down power of the pulsar, which is given by (see appendix)
\begin{equation}
L(t) = L_0 \left(1+\frac{t}{\tau_0}\right)^{-(n+1)/(n-1)}.
\label{eq:spinDown123}
\end{equation} 
The number of injected particles is assumed to be directly proportional to this spin-down power. We can thus change $L_0$ to inject more or fewer particles into the PWN. If more particles are injected into the PWN, the whole particle spectrum of the PWN will increase and thus also the radiation spectrum and vice versa (not shown). This change does not influence the shape of either the particle or the radiation spectrum. The same effect is seen when the value of the conversion efficiency ($\epsilon$) is changed (see Eq.~\eqref{normQ}). Another parameter is the characteristic spin-down timescale at birth ($\tau_{\rm{0}}$) given in Eq.~\eqref{eq:spinDown123} which characterises how fast the pulsar spins down. When this characteristic time is shorter, the pulsar spins down faster, resulting in fewer particles being injected into the PWN and thus the particle and radiation spectrum are both lower. The opposite happens when $\tau_0$ is longer. Furthermore, the braking index $n$ in Eq.~\eqref{eq:spinDown123} is usually set to 3 for rotating dipoles. If the braking index is decreased, the number of particles injected into the PWN increases due to the reduced spin-down of the pulsar (since both the index $(n+1)/(n-1)$ as well as $\tau_0$ change). Therefore, more particles are injected for longer periods into the PWN. Due to this, the particle and radiation spectrum will increase with a decreased $n$. These effects are similar to changing the normalisation of the injection spectrum.

\subsection{Soft-photon Fields}
\begin{figure}
\includegraphics[width=\columnwidth]{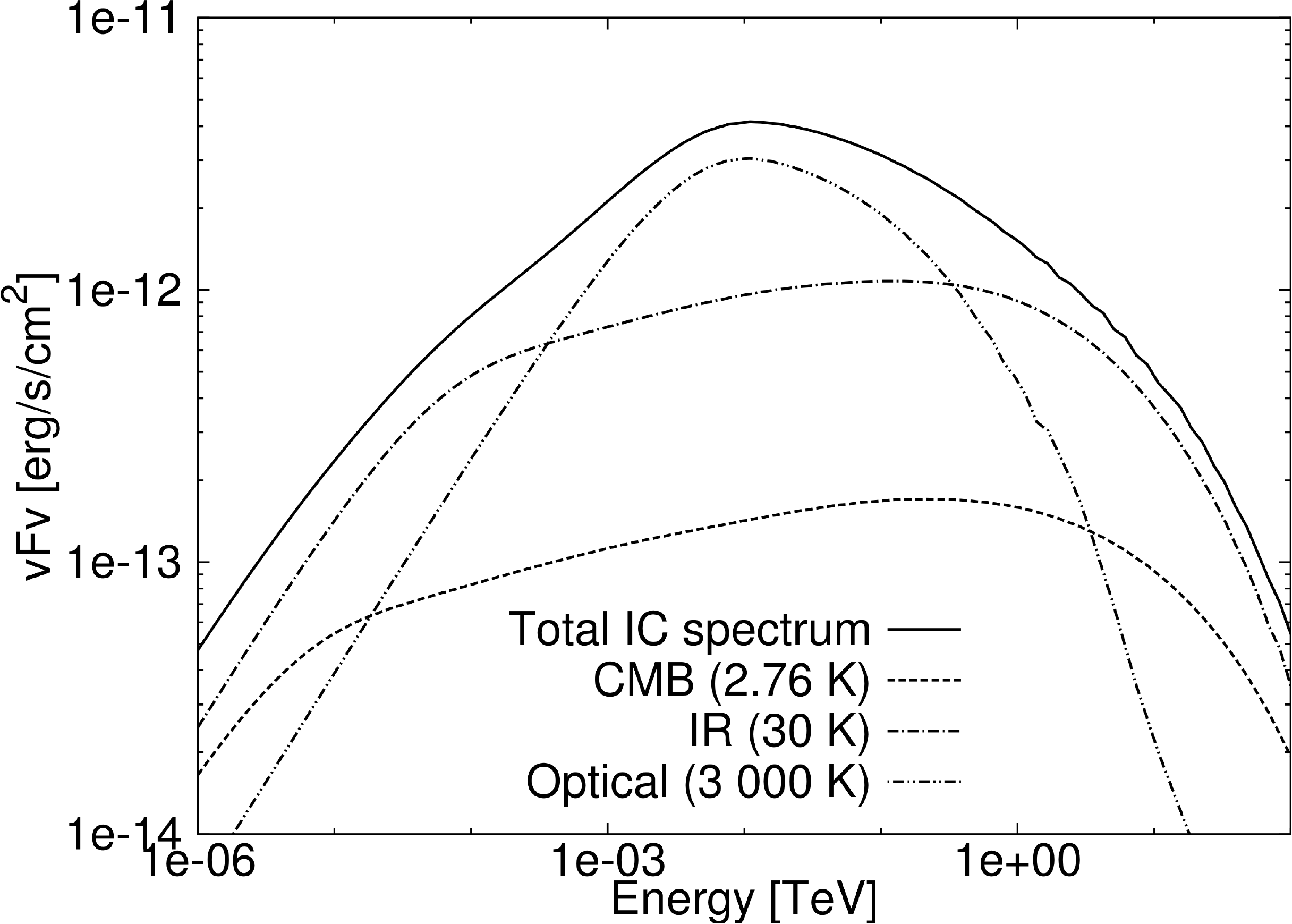}
\caption{\label{fig:SED_soft}IC spectrum for PWN~G0.9+0.1 showing the contribution of different soft-photon components in Table~\ref{tbl:G0.9_Torres}. The solid line is the total radiation, dashed line is the 2.76K CMB component, dashed-dotted line is the 30 K component, and the dashed-dot-dotted line shows the 3 000 K component.}
\end{figure}

Table~\ref{tbl:G0.9_Torres} shows the three different soft-photon components used to model the IC scattering in the PWN. These components can be turned on and off at will, and Figure~\ref{fig:SED_soft} shows the contribution of each of these components. The CMB target field produces a flat spectrum which causes the first small bump on the left hand side of the total IC flux component. The starlight at 3~000~K, with an energy density of 25 eV/cm$^3$, produces the highest peak and plays the largest role in the overall IC flux.
The effect of changes in the energy densities $u_l$ and the temperatures $T_l$ ($l=1,2,3$) of the soft-photon components can be understood as follows. For a single blackbody (BB), we have a spectral photon number density
\begin{equation}
  n_{\nu} = \frac{8\pi \nu^2}{c^3}\frac{1}{e^{h\nu/kT_l}-1}.
\end{equation}
For a given total photon energy density $u_l$ at a particular position in the PWN, we need a number $N_{\rm{BB}}$ of individual black bodies at a temperature $T_l$ to reach $u_l$, i.e., 
\begin{equation}
  N_{\rm{BB}} = \frac{u_l}{u_{\rm{BB}}(T_l)},
\end{equation}
with the energy density of a single BB
\begin{equation}
 u_{\rm{BB}}(T_l) = \int u_{\nu}d{\nu} = \int h \nu n_{\nu} d \nu \propto T_l^4
\end{equation}
the frequency-integrated energy density of a single BB and 
\begin{equation}
 \int n_{\nu} d \nu \propto T_l^3.
\end{equation}
Thus the IC flux from the PWN scales as (see also Eq. \eqref{eq:ICrad}, using $n_{\nu}$ instead of $n_{\varepsilon}$)
\begin{equation}
 \left( {\frac{dN}{dE}}\right)_{\rm{IC}} \propto  N_{\rm{BB}} \int n_{\nu} d\nu \propto \frac{u_l}{T_l}.
\end{equation}
Thus if the total energy density $u_l$ is increased or decreased, the IC radiation will also increase or decrease linearly. However, when the temperature is increased or decreased for a constant $u_l$, the IC flux scales in the opposite direction. This is due to the fact that when the temperature is increased, fewer photons are needed to reach the same energy density $u_l$ (since the average photon energy is now larger), leading to a lower normalisation for the cumulative BB spectrum.

\subsection{The Effect of Changing other Parameters}
The effects of changing most of the major parameters have been described, but the following are also free parameters worth noting. The free parameters $\alpha_1$ and $\alpha_2$ will influence the slopes and the normalisation of the particle and radiation spectrum. Lastly, the flux from the PWN at Earth scales as $1/d^2$ and the sizes of the spatial bins are also linearly dependent on $d$ (influencing the diffusion and convection timescales for each zone), but the latter is a small effect on the emitted SED.

\section{Spatially-dependent Results from Our PWN Model}
\label{sec:Space}
In the previous sections we showed the total particle spectrum and SED predicted by the code for different parameter choices. These calculations, however, were not the main aim of the code that we have developed, as we are especially interested in the spatial dependence of the radiation from the PWN. In this section we will study the effects that changing some of the parameters have on the energy-dependent size of the PWN.

\subsection{Effects of Changes in the Diffusion Coefficient and Bulk Particle Motion on the PWN's Morphology}\label{sec:morpho}
The diffusion coefficient contains two free parameters, which can be seen in Eq.~\eqref{eq:kappa123}. Here we consider the effects of changing the parameters $\kappa_0$ and $q$ (for Bohm diffusion, $q=1$), with $E'_0$ set to 1 TeV (changing $E'_0$ is similar to changing $\kappa_0$). We can now increase or decrease the value of $\kappa_0$ (assuming it is not linked with the magnitude of the $B$-field) and thus change the normalisation of the diffusion coefficient. We can also change $q$ which has an influence on the energy dependence of the diffusion coefficient:

\begin{figure}
\includegraphics[width=\columnwidth]{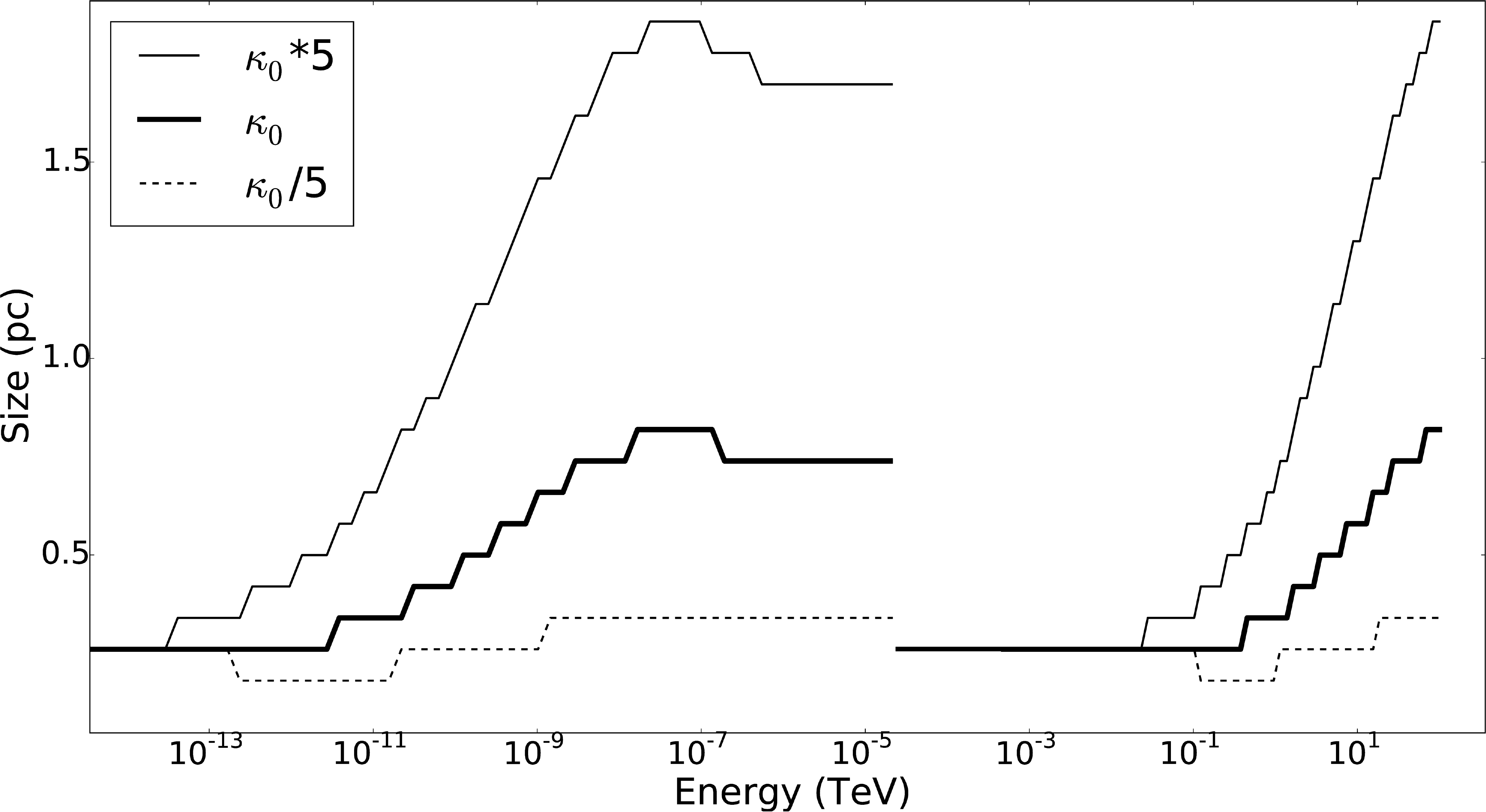}
\caption{\label{fig:SvE_k}Size of the PWN as a function of energy when the normalisation constant of the diffusion coefficient is changed.}
\end{figure}

Figure~\ref{fig:SvE_k} shows how the size of the PWN changes with energy for three different scenarios. The thin solid lines indicate 5$\kappa_0$, the thick solid lines indicate $\kappa_0$, and the dashed lines indicate $\kappa_0 /$5. The left graphs show SR and the right graphs IC emission. For this set of scenarios the size of the PWN increases with increased energy. In the first two scenarios, diffusion dominates the particle transport and causes the high-energy particles to diffuse outward faster than low-energy particles, filling up the outer zones and resulting in a larger size for the PWN at high energies. This effect is larger for high-energy particles due to the energy dependence of the diffusion coefficient ($q > 0$). For $\kappa_0/$5, we see that the effect is not as pronounced. Here the diffusion coefficient is so small that the SR energy loss rate starts to dominate diffusion. The particles therefore ``burn off" or expend their energy before they can reach the outer zones (cooling therefore dominates). Changes to $q$ have similar effects on the SED than changes to $\kappa_0$ but are more pronounced at higher energies.

\begin{figure}
\includegraphics[width=\columnwidth]{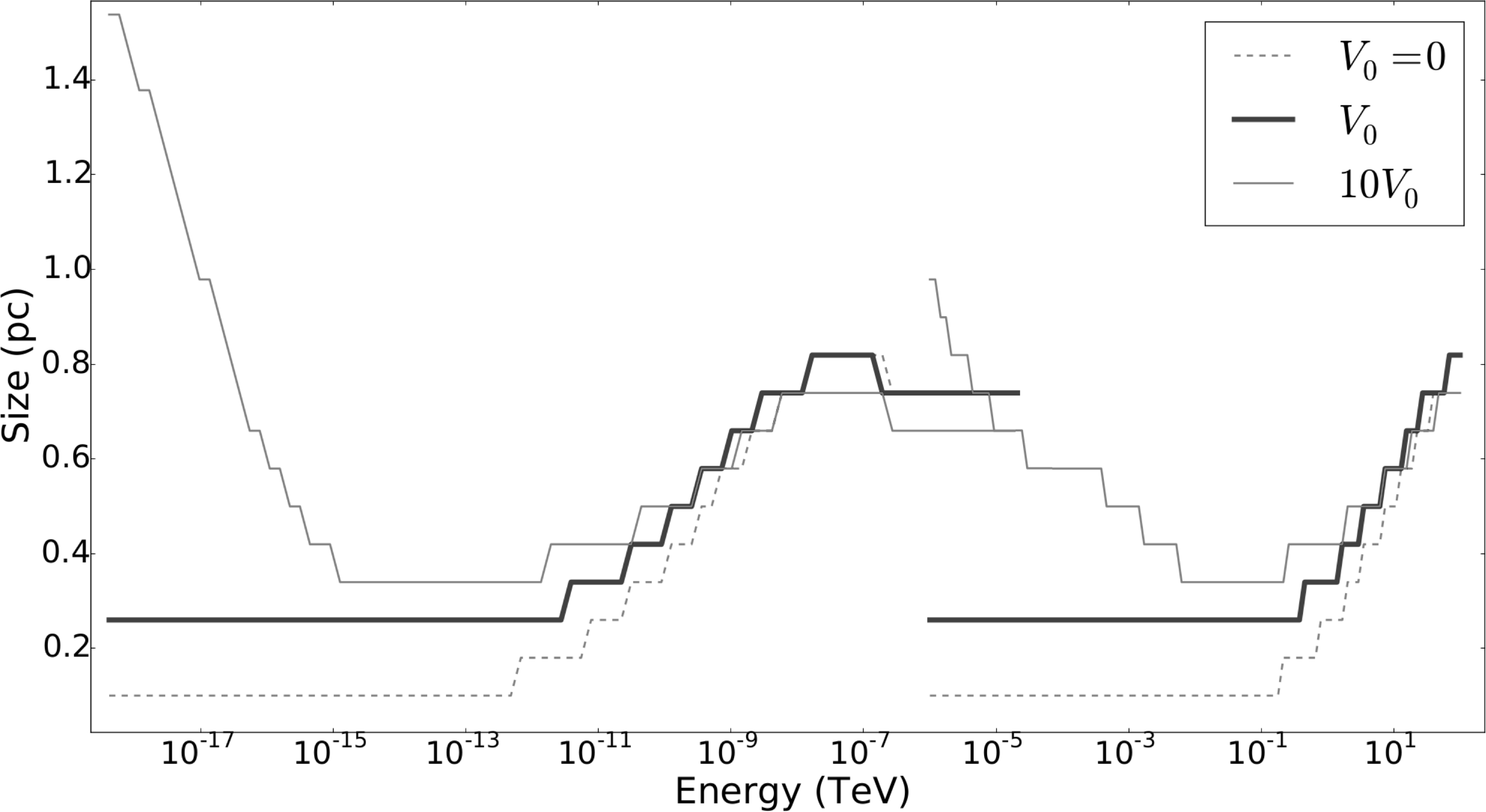}
\caption{\label{fig:SvE_V}Size of the PWN as a function of energy for different normalisations of the bulk particle speed for the model parameters given in Table~\ref{tbl:G0.9_Torres}.}
\end{figure}

\begin{figure}
\includegraphics[width=\columnwidth]{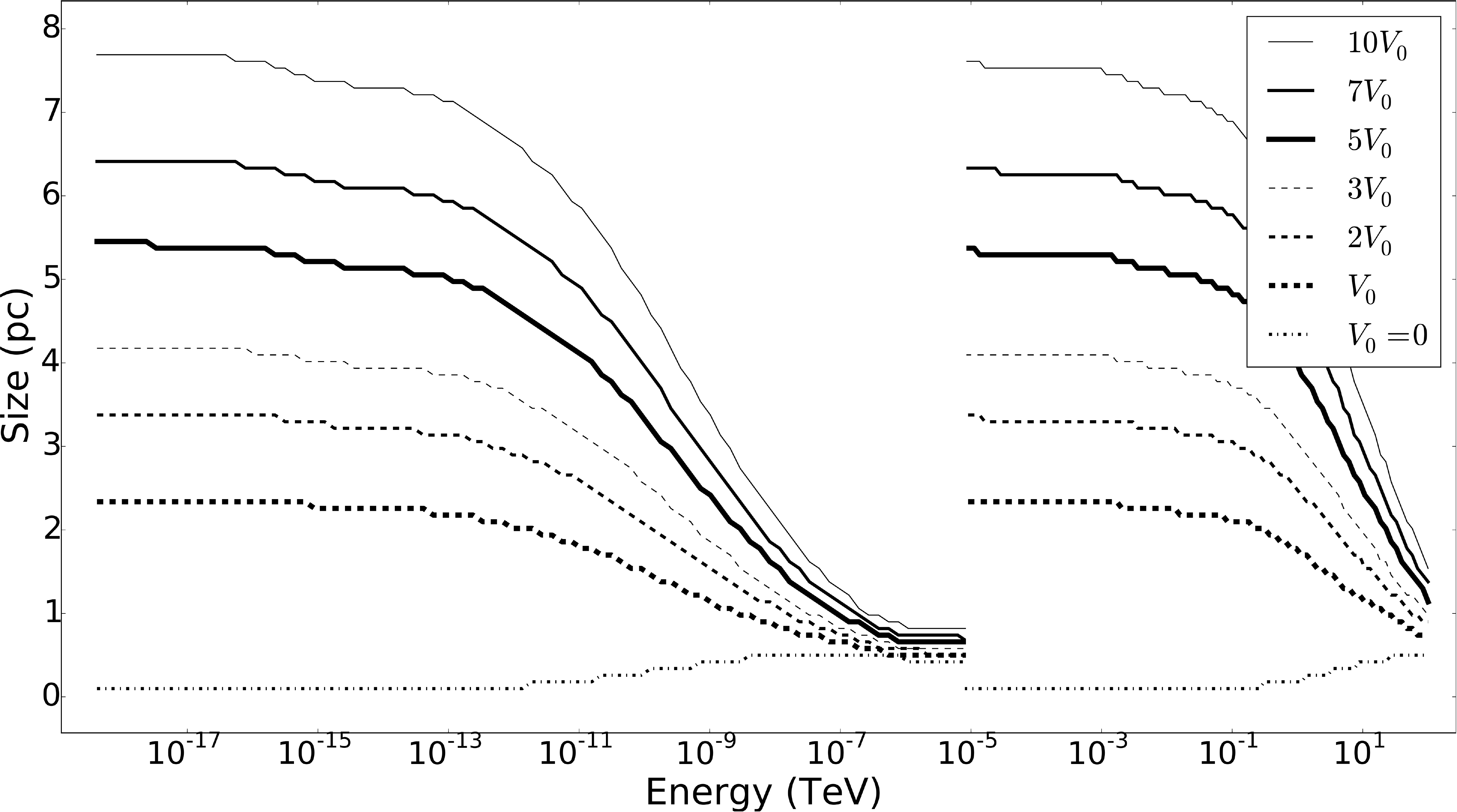}
\caption{\label{fig:SvE_V_Carlo}Size of the PWN as a function of energy for different normalisations of the bulk particle speed for the model parameters given in Table~\ref{tbl:J1356_Carlo}.}
\end{figure}

Next we studied the effect of varying the bulk motion on the energy-dependent size of the PWN by varying $V_0$ for two different cases: the first as seen in Figure~\ref{fig:SvE_V} is for the model parameters given in Table~\ref{tbl:G0.9_Torres}, while the second as seen in Figure~\ref{fig:SvE_V_Carlo} is for the parameters given in Table~\ref{tbl:J1356_Carlo}. If we consider $V_0 = 0$ (Figure~\ref{fig:SvE_V}), indicated by the dashed line, we find that the size of the PWN is determined by the energy-dependent diffusion and therefore the size increases with increasing energy. Adding a bulk flow to the code (e.g., non-zero $V_0$, thick solid line) increases the size of the PWN irrespective of the energy of the particles. However, for a very large bulk flow speed (e.g, $10V_0$, thin solid line), the radio size is  significantly larger than the X-ray size. This is due to the energy-dependence of the SR energy losses which dominate at higher energies, thereby reducing the lifetime of these X-ray-emitting particles and resulting in a smaller source compared to the radio. In this first case $\alpha_{\rm{V}} = 1$, which is non-physical as mentioned in the discussion following Eq.~\eqref{eq:ad_parmtr2}. The bulk flow speed becomes so large in the outer zones that particle escape becomes significant. Therefore, if the normalisation is increased beyond $10V_0$, the radio source size in fact starts to decrease. Next we do a similar study by using the more physical set of parameters given in Table~\ref{tbl:J1356_Carlo}, where $\alpha_{\rm{V}} = -1$. Figure~\ref{fig:SvE_V_Carlo} shows the effect of changes to the normalisation of the bulk motion of particles.
Again, if $V_0 = 0$ (dashed-dotted line, same line as in Figure~\ref{fig:SvE_V}), the PWN has a smaller size at lower energies than at higher energies. The size increases monotonically with $V_0$. At the lower energies convection dominates the radiative energy losses and therefore the particles have a very long lifetime, allowing them to diffuse to the outer zones and still be able to radiate, resulting in a large source size. In contrast to this, at high energies, the SR losses dominate the convection, resulting in a very short lifetime for the high-energy particles, therefore these particles radiate all their energy before they have time to convect to the outer zones. This leads to a relatively smaller X-ray source size.

\subsection{Different Cases of $\alpha_{\rm{V}}$ and $\alpha_{\rm{B}}$}\label{sec:alph_Valph_B}
In Eq.~\eqref{V_profile} and Eq.~\eqref{B_Field} we assumed that the $B$-field may have a spatial and time dependence and that the bulk motion only has a spatial dependence. In this section the effects of different spatial dependencies for $B(r,t)$ and $V(r)$ are studied. Note that we have assumed the diffusion coefficient to be spatially independent throughout this work. However, since we are now considering the spatial dependence of the $B$-field in this paragraph, and $\kappa \propto 1/B(r,t)$, this assumption is technically violated here. The effect is small when the divergence of $\vec{\kappa}$ is small, which we will assume to be the case in this section. This spatial dependence of the diffusion coefficient can be implemented in future by adding another convective term to the transport equation. 

From Eq.~\eqref{eq:a_v+a_b=-1}, the following relationship is assumed to hold: $\alpha_{\rm{V}} + \alpha_{\rm{B}} = -1$. For this section the time dependence of the $B$-field is kept unchanged, with $\beta_{\rm{B}} =-1.3$, and four different scenarios for $\alpha_{\rm{B}}$ and $\alpha_{\rm{V}}$ are shown. Here the first situation is the same as \cite{Torres2014}, with $\alpha_{\rm{B}} = 0$ and $\alpha_{\rm{V}} = 1$, We also considered the following three situations: $\alpha_{\rm{B}} = 0$ and $\alpha_{\rm{V}} = -1$, $\alpha_{\rm{B}} = -0.5$ and $\alpha_{\rm{V}} = -0.5$, and $\alpha_{\rm{B}} = -1$ and $\alpha_{\rm{V}} = 0$. These three situations all comply with Eq.~\eqref{eq:a_v+a_b=-1}, with the $B$-field kept constant in the first spatial zone. The $B$-field was limited to a maximum value, as the parametrisation resulted in the $B$-field growing infinitely large during the early epochs of the PWN's lifespan.

\begin{figure}
\includegraphics[width=\columnwidth]{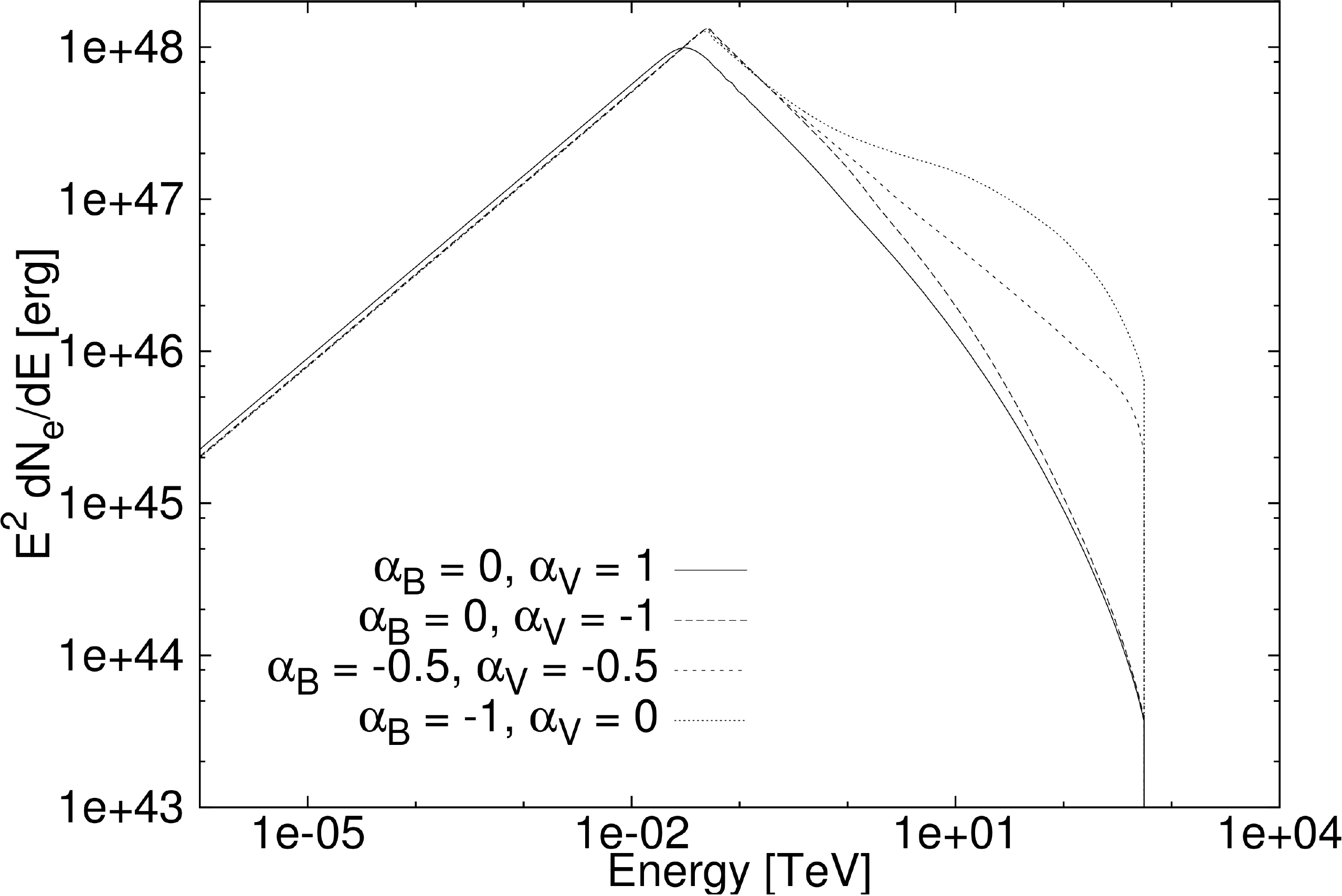}
\caption{\label{fig:Par_Change_alp}Particle spectrum for PWN G0.9+0.1 with a change in the parametrised $B$-field and bulk particle motion.}
\end{figure}

\begin{figure}
\includegraphics[width=\columnwidth]{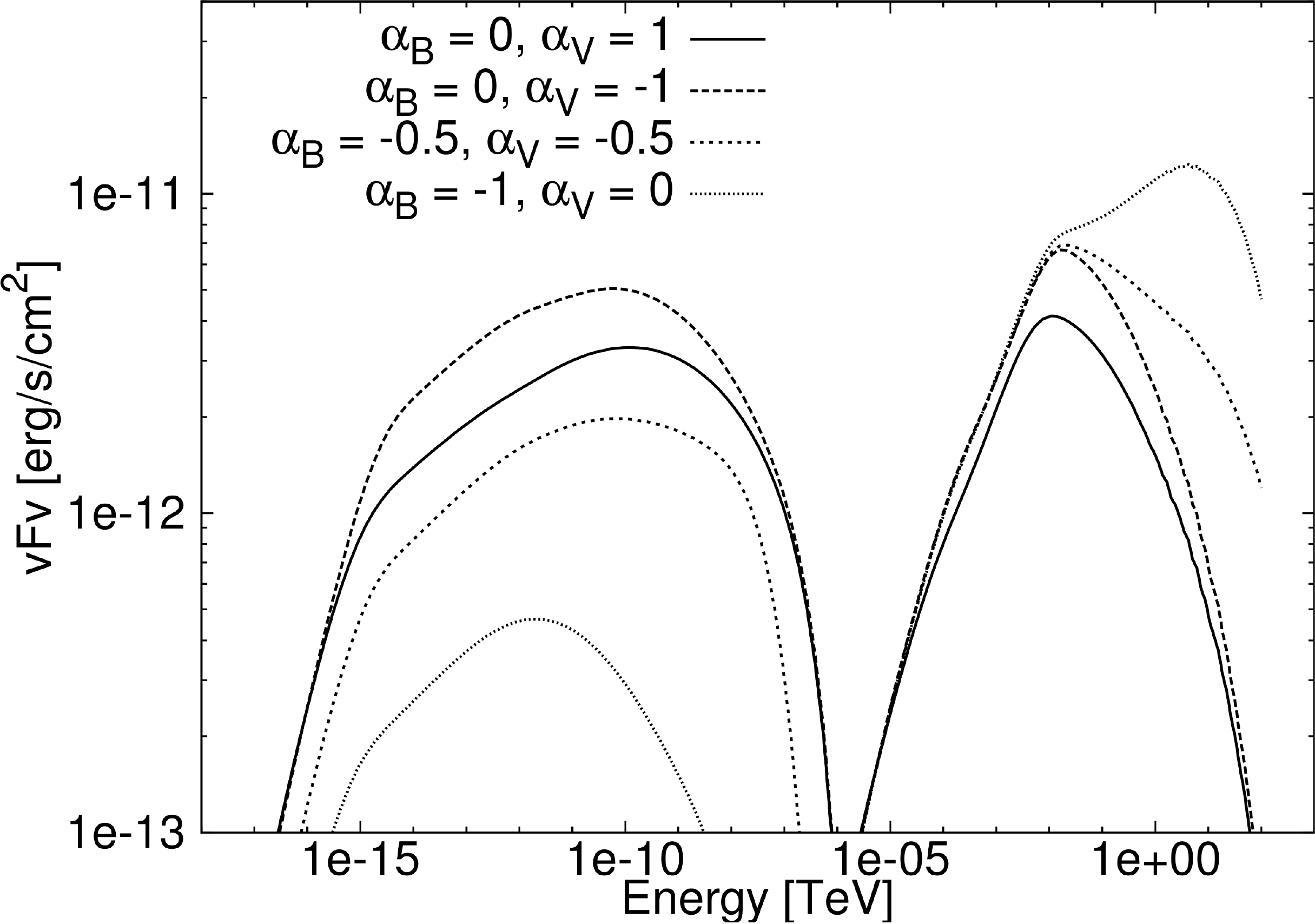}
\caption{\label{fig:SED_Change_alp}SED for PWN G0.9+0.1 with a change in the parametrised $B$-field and bulk particle motion.}
\end{figure}

In Figure~\ref{fig:Par_Change_alp} the particle spectrum is shown for the four different scenarios, with the solid line showing the result for $\alpha_{\rm{B}} = 0$ and $\alpha_{\rm{V}} = 1$ as is effectively assumed by \cite{Torres2014}. In this case the $B$-field is spatially constant over the entire PWN, but the bulk speed increases linearly with $r$. The particles move extremely fast as they propagate farther from the centre of the PWN. They therefore lose more energy due to adiabatic energy losses relative to the other cases. Thus the solid line is lower than the other situations and the peak of the spectrum is also shifted to the left.

We can see from both Figures~\ref{fig:Par_Change_alp} and \ref{fig:SED_Change_alp} that changes to the $B$-field have a more profound impact on the particle spectrum and SED than changes to the radially-dependent speed. If the spatial dependence of the $B$-field changes from $\alpha_{\rm{B}}=0$ to $\alpha_{\rm{B}}=-0.5$ and $\alpha_{\rm{B}}=-1$, the $B$-field is first constant over all space and then decreases as $r^{-0.5}$ and finally it reduces rapidly as $r^{-1}$. The effect of this can be seen in the particle spectrum as the number of high-energy particles increases for a decreased $B$-field and hence a lower SR loss rate. This effect is emphasised in the situation where $\alpha_{\rm{B}} = -1$, resulting in a very small $B$-field at the outer edges of the PWN. This can also be seen in the radiation spectrum in Figure~\ref{fig:SED_Change_alp} where a decreased $B$-field results in reduced radiation in the SR band (since $\dot{E}_{\rm{SR}} \propto B^2$), and the increased radiation in the IC band is due to more particles being present at those energies. This increase in the high-energy particles is quite large for $\alpha_{\rm{B}} = -1$, though (possibly indicating a violation of our assumption that the divergence of $\vec{\kappa}$ is small in this case). We note that our model currently does not take into account the fact that the cutoff energy due to particle escape ($E_{\rm{max}}$) should also be a function of the $B$-field. This is because in reality $\sigma \propto B^2$ (we have assumed $\sigma$ to be constant), and therefore $E_{\rm{max}} \sim \sqrt{B^2/(1+B^2)}$, which will have the effect that if the $B$-field is reduced, $\sigma$ and therefore $E_{\rm{max}}$ will decrease. This may cause the high-energy particles to be cut off at lower energies as the $B$-field decreases due to more efficient particle escape, and therefore the build up of high-energy particles may be partially removed (we say `partially' since the Larmor radius of the most energetic particles in the outer zones is still smaller than the PWN size by a factor of a few, inhibiting efficient escape of particles from the PWN). The question of particle escape may also be addressed by refining our outer boundary condition in future.

\begin{figure}
\includegraphics[width=\columnwidth]{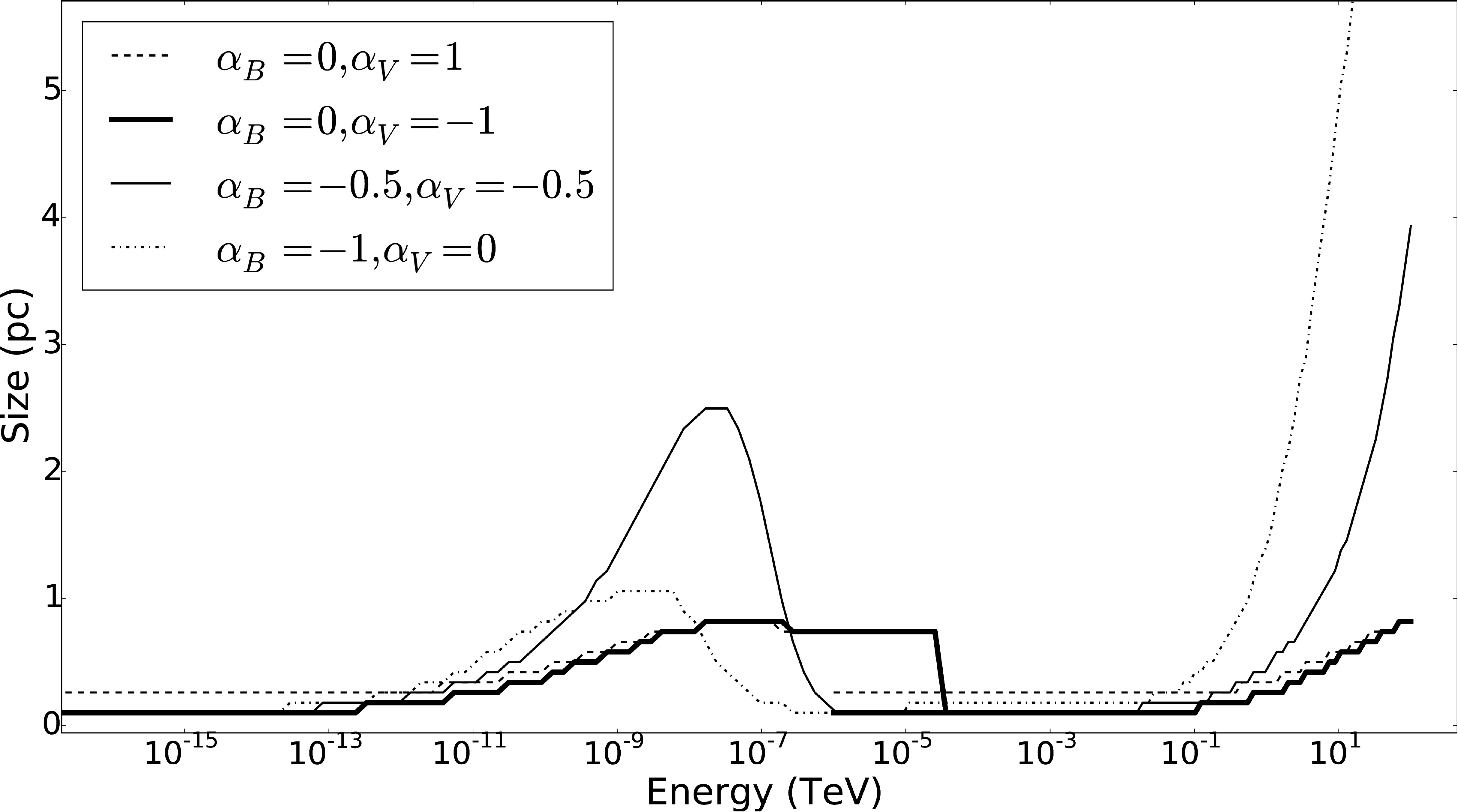}
\caption{\label{fig:SvE_alpB}Size of PWN G0.9+0.1 as a function of energy for changes in $\alpha_{\rm{B}}$ and $\alpha_{\rm{V}}$.}
\end{figure}

From Figure~\ref{fig:SvE_alpB} we can see that in scenario one (dashed line, $\alpha_{\rm{B}} = 0$ and $\alpha_{\rm{V}} = 1$) the PWN size for low energies is always larger than for all the other scenarios. This is due to the bulk speed being directly proportional to $r$ in this case, resulting in the particles moving faster as they move farther out from the centre of the PWN. This will result in the outer zones filling up with particles, while not escaping. This may point to our outer boundary that was chosen to be much larger than the radius of the PWN ($r_{\rm{max}} \gg R_{\rm{PWN}}$). For scenario two (thick solid line, $\alpha_{\rm{B}} = 0$ and $\alpha_{\rm{V}} = -1$), the size of the PWN  at low energies follows the same pattern as for both low-energy and high-energy photons, since the energy-dependent diffusion now dominates convection. At lower energies, we see that PWN is smaller than in scenario one, as the speed is now proportional to $r^{-1}$, which results in a slower bulk motion and thus fewer low-energy particles moving to the outer zones. In scenario three, (thin solid line, $\alpha_{\rm{B}} = -0.5$ and $\alpha_{\rm{V}} = -0.5$), and four (dotted line, $\alpha_{\rm{B}} = -1$ and $\alpha_{\rm{V}} = 0$) the $B$-field has a spatial dependence, reducing as one moves farther away from the centre of the PWN. This reduced $B$-field will lead to increased diffusion and decreased SR losses as mentioned in the first part of this section. For these two scenarios the dependence of the bulk motion on radius is weaker and therefore diffusion dominates the particle transport. Once again we can see the energy dependence of the diffusion, since the PWN is initially smaller and then increases for higher energies. At very high energies, the PWN size becomes very large, which is not the case for the SR component. The first is due to the pile up of high-energy particles (leading to substantially increased IC emission, Figure~\ref{fig:SED_Change_alp}), while the second is due to the fact that SR is severely inhibited for the very low $B$-field.

\subsection{Size versus Energy Fits}
\begin{figure}
\includegraphics[width=\columnwidth]{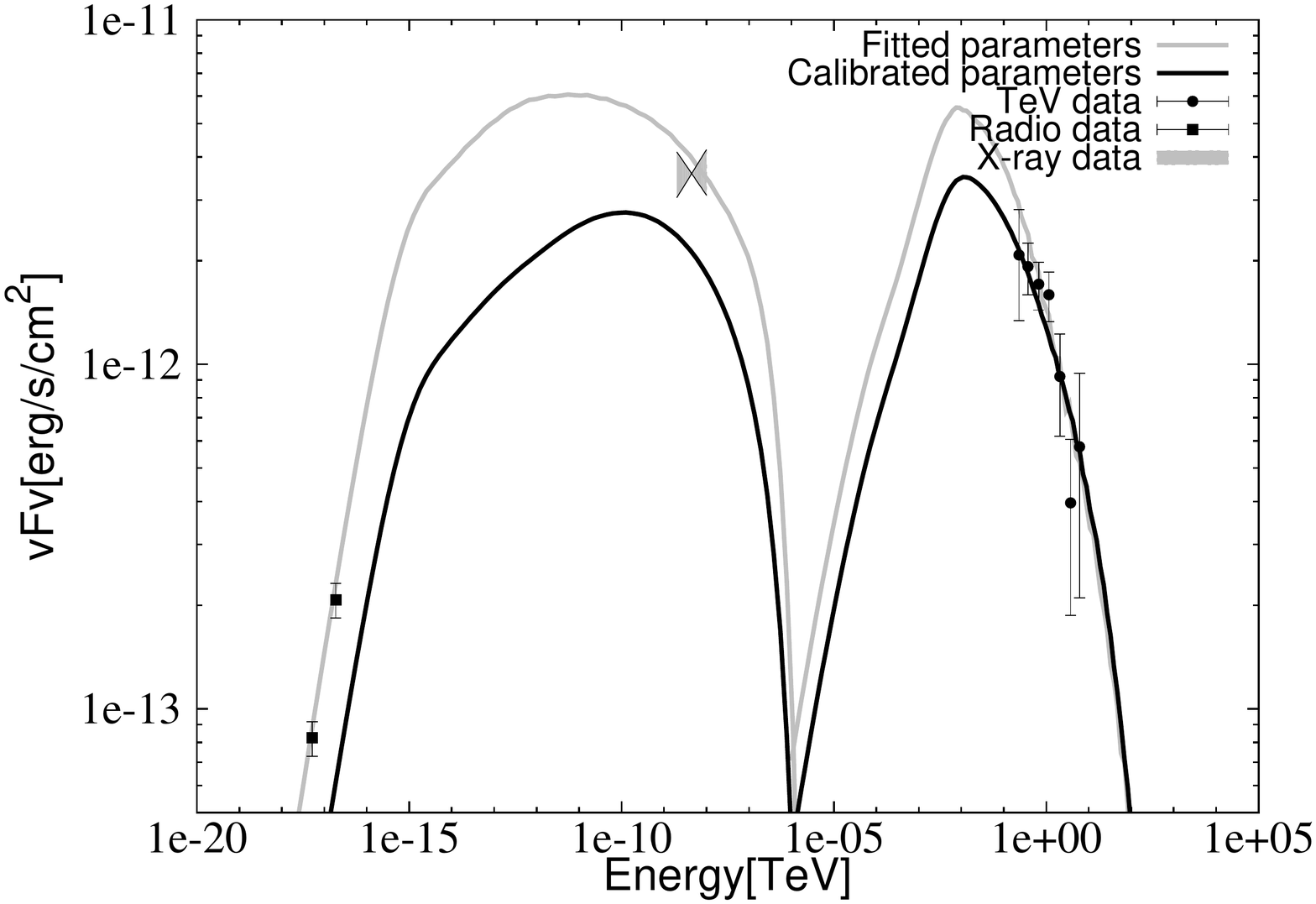}
\caption{\label{fig:SED_combine}SED for PWN G0.9+0.1 for the parameters used by \citet{Torres2014} (Table~\ref{tbl:G0.9_Torres}) and the fitted parameters as in Table \ref{tbl:J1356_Carlo}. The radio \citep{HelfandB1987}, X-ray \citep{Porquet2003} and $\gamma$-ray data \citep{G0.9+0.1_HESS} are also shown.}
\end{figure}

\begin{figure}
\includegraphics[width=\columnwidth]{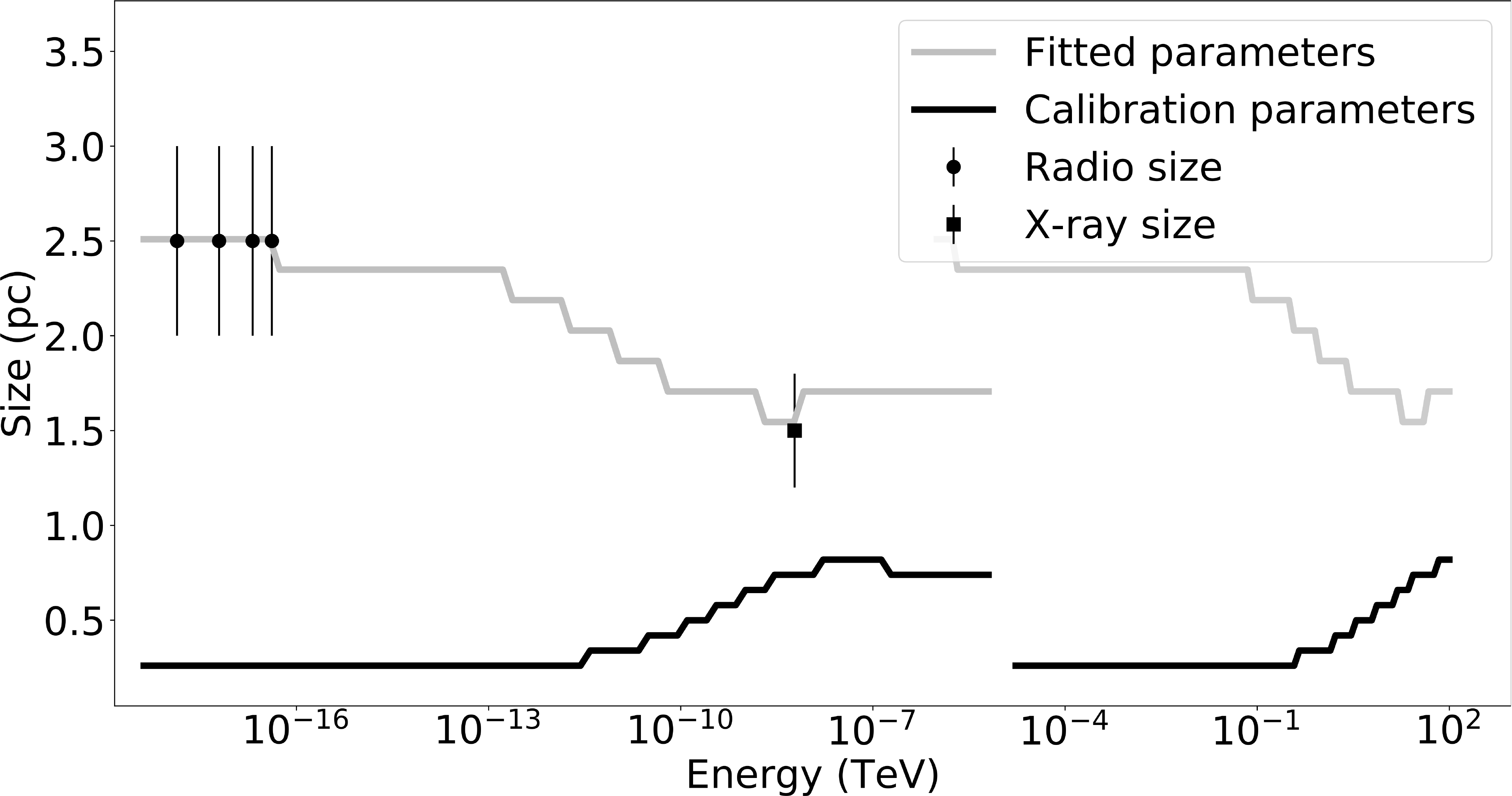}
\caption{\label{fig:SvE_combine}Size of the PWN as a function of energy for the calibration parameters in Table \ref{tbl:G0.9_Torres} and the fitted parameters in Table \ref{tbl:J1356_Carlo}. The observed radio \citep{Dubner2008} and X-ray sizes \citep{Porquet2003} are also indicated.}
\end{figure}

Figure \ref{fig:SED_combine} and \ref{fig:SvE_combine} show the radiation spectrum and the size versus energy graphs for PWN G0.9+0.1 for the calibration parameters (black lines) as in Table \ref{tbl:G0.9_Torres} modelled by \cite{Torres2014}, with the dots indicating the estimated radio and the square the estimated X-ray size\footnote{We directly infer the source sizes for the radio and X-rays bands from the respective images in the original papers. This procedure may be somewhat arbitrary and prone to error, and also dependent on the presentational choices or assumptions made in the original papers. The best way to determine these sizes would be to redo the data analysis and infer them in a systematic way. However, this is beyond the scope of the current modelling paper. Furthermore, we note that our model is spherically symmetric, while the data indicate that the source is not. To account for these uncertainties (i.e., source asymmetry and actual energy-dependent source size), we specified a sizable error on our estimated values.}. The upper limit on the predicted TeV size is 10.4 pc, i.e., we use the point spread function of the H.E.S.S. telescope (not shown). The radio data are from \cite{HelfandB1987} and \cite{Dubner2008}, the X-ray data are from \cite{Porquet2003}, and the TeV data from \cite{G0.9+0.1_HESS}. The model provides reasonable fits to the spectral radio, X-ray, and TeV data, however, it is clear that the predicted size of the PWN does not fit the data at all. After adjusting some parameters, we found a better fit and this can also be seen in Figure \ref{fig:SED_combine} and \ref{fig:SvE_combine} (grey lines). Table \ref{tbl:J1356_Carlo} shows the new parameters used for this fit.

\begin{table}
\begin{center}
\caption{Modified parameters for PWN G0.9+0.1 for fitting the SED as well as the energy-dependent size of the PWN.\label{tbl:J1356_Carlo}}
\resizebox{\columnwidth}{!}{
\begin{tabular}{crrr}
\hline
Model Parameter & Symbol & Value & \cite{Torres2014}\\
\hline
Present-day $B$-field & $B(t_{\rm{age}})$ & 15.98.0 $\mu \rm{G}$ & 14.0 $\mu \rm{G}$\\
Age of the PWN & $t_{\rm{age}}$ & $3~227~\rm{yr}$ &  $2~000~\rm{yr}$ \\
Initial spin-down power ($10^{38}\rm{erg}$ $\rm{s^{-1}}$)& $L_0$ & 1.44  & 1.0\\ 
$B$-field parameter & $\alpha_{\rm{B}}$ & 0.0 & 0\\
$B$-field parameter & $\beta_{\rm{B}}$ & $-$1.0 & $-1.3$\\
$V$ parameter & $\alpha_{\rm{V}}$ & $-1.0$ & $1.0$\\
Particle bulk motion & $V_0$ & $0.0615~c$ & $1.63 \times 10^{-4} c$\\
Diffusion & $\kappa_0$ & $3.356$ & $1.0$\\
\hline
\end{tabular}
}
\end{center}
\end{table}

The process of finding a better fit to the both the SED of the PWN and the energy-dependent size was facilitated by our prior parameter study. The only way in which we could increase the size of the PWN at lower energies was to increase the bulk speed of the particles. This, however, increased the adiabatic energy losses, and resulted in a lower radiation spectrum. This was then countered by increasing the age of the PWN (which effectively leads to an increase in $L_0$). The bulk speed of the particles had to be increased substantially to fit the data, but given the large errors on the size of the PWN in the radio band,  we could still fit the data with a bulk speed as small as $0.073c$. The profile trends for the $B$-field as well as the bulk speed of the particles were also changed. To increase the size of the PWN further we also increased the normalisation of the diffusion coefficient of the particles. This is not a bad assumption as the diffusion was originally modelled to be Bohm-type diffusion, which is a very slow perpendicular diffusion with respect to the $B$-field. All these changes produced the grey lines in Figure \ref{fig:SED_combine} and \ref{fig:SvE_combine}. Here we see that we have a good fit for the radio size, which according to data, does not change with energy and the model reproduces this trend as well as the trend where the size of the PWN decreases with increasing energy. This is a common feature of PWNe.

In a future paper a more robust statistical method may be used to find the best fit to this source's SED and energy-dependent size and to also investigate the parameter degeneracy.

\section{Conclusions} 
\label{sec:concl}
This study focused on modeling the evolution of PWNe, with the main aim being to create a spatially-dependent temporal code to model the radiation morphology of PWNe. We solved a Fokker-Planck-type transport equation to model the particle evolution inside a PWN, injecting a broken power-law particle spectrum and allowing this spectrum to evolve over time taking into account energy losses due to SR, IC scattering, and adiabatic cooling of the PWN due to expansion. We also took into account particle diffusion and convection in the form of a bulk particle motion. \textit{Our model is now able to not only fit the observed radiation spectra the PWN, but also yields results concerning the morphology of the PWN (i.e., it is able to reproduce the size of the PWN as a function of energy). Thus we can potentially derive stronger constraints on key quantities characterising the PWN.}

We calibrated the code by comparing it to results by two independent codes (\citealt{VdeJager2007} and \citealt{Torres2014}), using PWN G0.9+0.1 as calibration source. We found that our model was well calibrated. Our model is now able to not only fit the observed radiation spectra from the PWN, but also yields results concerning the morphology of the PWN (i.e., it is able to reproduce the size of the PWN as a function of energy). Thus we can potentially derive stronger constraints on key quantities characterising the PWN.

The spatio-temporal-energetic model we presented is a first approach to modelling PWNe for multiple spatial bins, thus there are a number of improvements that can be made. For example, the code currently has a problem with a build up of particles at high energies when the $B$-field decreases rapidly with radius. This is partially due to the fact that we chose a fixed $r_{\rm{max}}\gg R_{\rm{PWN}}$. We will revise this boundary condition in future. One way in which this could be refined is by using an MHD code to model the morphology of the PWN in more detail and to find a more realistic time-dependent radius of the PWN. This will allow us to use this radius as the outer boundary which will enable the particles to escape more efficiently from the PWN. One can also obtain more realistic spatial and time dependencies for the $B$-field and bulk flow speeds using an MHD code. This will yield refined SR and adiabatic losses and convection. Furthermore, treating $\sigma$ as being dependent on the $B$-field will aid by lowering the maximum energy of particles that are contained within the PWN. The code should also be generalised in future to handle a spatially-dependent diffusion coefficient by adding another convective term to the transport equation.

In future we will perform a population study to investigate currently known trends, e.g., the X-ray luminosity that correlates with the pulsar spin-down luminosity and its anti-correlation with the characteristic age of the pulsar. We could also probe other trends, e.g., investigate whether there is a correlation between the TeV surface brightness of the PWN and the spin-down luminosity of the pulsar \citep{PWN_pop2017}, as well as the surface brightness versus age. Some follow-up projects or refinements to the model are as follows. The code currently assumes spherical symmetry. This can be revised by expanding the model to two or three spatial dimensions. One could also add anisotropic effects such as considering distinct equatorial and polar outflows (injection) of particles.  Some older PWNe are offset from the pulsar, revealing a bullet shape. 
This is either due to an inhomogeneity in the interstellar medium (ISM) in which the PWN expands causing an asymmetric reverse shock and thus an offset PWN, or to the pulsar receiving some kick velocity at birth, thus moving away from the PWN centre. The radiation peaks at the pulsar position, thus also causing the bullet shape. These effects could be added to the model to simulate a more realistic situation. The code is currently only applicable to young PWNe. This should be addressed so that all ages of PWNe can be modelled, e.g., by including a more complex parametrisation of the $B$-field and adding the effect of an asymmetric reverse shock to the code. 

The CTA will reveal more sources and more information regarding the morphology of PWNe due to its improved sensitivity and angular resolution. This will necessitate the continued development, application, and refinement of spatially-dependent PWN codes such as the one discussed here.

\section*{Acknowledgements}
We gratefully acknowledge fruitful discussions with Andreas Kopp, Harm Moraal, and Stefan Ferreira. This work is based on the research supported wholly / in part by the National Research Foundation (NRF) of South Africa (Grant Numbers 87613, 90822, 92860, 93278, and 99072). The Grantholder acknowledges that opinions, findings and conclusions or recommendations expressed in any publication generated by the NRF supported research is that of the author(s), and that the NRF accepts no liability whatsoever in this regard.

\label{Bibliography}
\bibliographystyle{mnras}  
\bibliography{Bibliography}

\section{Appendix} \label{ap:1}
We derive expressions for $L(t)$ and $L_0(\tau_0)$ to show that $\tau_0 = \tau_{\rm{c}} - t_{\rm{age}}$. We make two assumptions: the first is that the $B$-field of the pulsar does not decay over short time scales, i.e., $\dot{P}P^{n-2} = \dot{P}_0P_0^{n-2}$ \citep[e.g.,][]{VdeJager2007} and the second is a braking law of the form $\dot{\Omega} = k \Omega^n$ \citep[e.g.,][]{Pacini1973, Rees1974, Gaensler06}.

The spin-down luminosity $L(t)$ of the pulsar can be constructed by using the second assumption and the following definition $L(t) = I \Omega \dot{\Omega}$, thus $L(t) = kI \Omega^{n+1}$.
We can integrate $\dot{\Omega}$ to find
\begin{equation}
 \int_{\Omega_0}^{\Omega} \Omega^{-n}d\Omega=\int_0^tdt,
\end{equation}
thus 
\begin{equation}
 \frac{1}{1-n}\left(\Omega^{1-n}-\Omega_0^{1-n}\right) = kt
\end{equation}
leaving us with 
\begin{equation}
\label{eq:Omega}
 \Omega = \left(\frac{1}{(1-n)kt+\Omega_0^{1-n}}\right)^{\frac{1}{n-1}}.
\end{equation}
Now we can obtain $L(t)$ by replacing the $\Omega$ with Eq. \eqref{eq:Omega}. Thus
\begin{equation}
 L(t) = kI\left(\frac{1}{(1-n)kt+\Omega_0^{1-n}}  \right)^{\frac{n+1}{n-1}}.
\end{equation}
We set $\beta = (n+1)/(n-1)$ and do some manipulation to find
\begin{equation}
 L(t) = kI\Omega_0^{n+1}\left( 1+\frac{(1-n)kt}{\Omega_0^{1-n}} \right)^{-\beta}.
\end{equation}
We know from the definition of $L(t)$ that $L_0(t) = kI\Omega_0^{n+1}$ (assuming a constant value for $I$ and $k$) and also that $(1-n)k/\Omega_0^{1-n} = (1-n)\dot{\Omega}_0/\Omega_0 = 1/\tau_0$ and thus 
\begin{equation}
\label{eq:L_t_fo_tau_023}
 L(t) = \frac{L_0}{\left( 1+\frac{t}{\tau_0} \right)^{\beta}}.
\end{equation}

In this first part we have shown how the spin-down luminosity is derived from the second assumption. When substituting $t=t_{\rm{age}}$ and $L_{\rm{age}} = L(t_{\rm{age}}) = 4\pi^2I\dot{P}/P^3$, we find a first expression for $L_0$:
\begin{equation}
\label{eq:L_t_fo_tau_0}
 L_0 = L_{\rm{age}}\left( 1+\frac{t_{\rm{age}}}{\tau_0} \right)^{\beta}.
\end{equation}
We will now obtain another expression for $L_0(\tau_0)$ using the first assumption $\dot{P}P^{n-2} = \dot{P}_0P_0^{n-2} = K$, with $K$ a constant. We rewrite this assumption as:
\begin{equation}
 P_0 = \left(\frac{K}{\dot{P}_0}\right)^{\frac{1}{n-2}}.
\end{equation}
Since $L_0 = 4\pi^2I\dot{P}_0/P_0^3$
\begin{equation}
 L_0=\frac{4\pi^2I\dot{P}_0}{\left(\frac{K}{\dot{P}_0} \right)^{3/(n-2)}}.
\end{equation}
Following some manipulations we find
\begin{equation}
\label{eq:L_0_fo_P0}
 L_0 = \frac{4\pi^2I}{K^{3/(n-2)}} \cdot \dot{P}_0^{\frac{n+1}{n-2}}.
\end{equation}
We can also find $\dot{P}_0$ as a function of $\tau_0$ by using the definition for the birth characteristic age of the pulsar given by $\tau_0~=~P_0/(n-1)\dot{P}_0$. Thus we have
\begin{equation}
 \tau_0 = \frac{(K/\dot{P}_0)^{1/(n-2)}}{(n-1)\dot{P}_0},
\end{equation}
and once again we solve for $\dot{P}_0$, leaving us with
\begin{equation}
\label{eq:P_dot_0}
 \dot{P}_0 = \left( \frac{K^{1/(n-2)}}{(n-1)\tau_0} \right)^{\frac{n-2}{n-1}}.
\end{equation}
We can now substitute Eq. \eqref{eq:P_dot_0} into Eq \eqref{eq:L_0_fo_P0} to find $L_0$ as a function of $\tau_0$. We are thus left with
\begin{equation}
 L_0 = \left( \frac{4\pi^2I}{K^{3/(n-2)}} \right)\left( \left[ \frac{K^{1/(n-2)}}{(n-1)\tau_0} \right]^{\frac{n-2}{n-1}} \right)^{\frac{n+1}{n-2}},
\end{equation}
resulting in
\begin{equation}
 L_0 = 4\pi^2I K^{-2/(n-1)}\left( \frac{1}{(n-1)\tau_0} \right)^{\frac{n+1}{n-1}}.
\end{equation}
By substituting the constant $K=\dot{P}P^{n-2}$ back we find
\begin{equation}
 L_0 = 4\pi^2I \dot{P}^{\frac{-2}{(n-1)}} P^{\frac{-2(n-2)}{(n-1)}} \left( \frac{1}{(n-1)\tau_0} \right)^{\frac{n+1}{n-1}},
\end{equation}
and by using the definition for the current spin-down luminosity $L_{\rm{age}}=4\pi^2I\dot{P}/P^3$ we find
\begin{equation}
 L_0 = L_{\rm{age}} \frac{P^3}{\dot{P}} \dot{P}^{\frac{-2}{(n-1)}} P^{\frac{-2(n-2)}{(n-1)}} \left( \frac{1}{(n-1)\tau_0} \right)^{\frac{n+1}{n-1}}.
\end{equation}
Upon simplification we find
\begin{equation}
 L_0 = L_{\rm{age}}\left( \frac{P}{\dot{P}(n-1)\tau_0} \right)^{\frac{n+1}{n-1}}.
\end{equation}
We can simplify this further by using the definition for the characteristic age of the pulsar, thus
\begin{equation}
\label{eq:L_t_fo_tau_02}
 L_0 = L_{\rm{age}}\left( \frac{\tau_c}{\tau_0} \right)^{\beta}.
\end{equation}

We now have two forms for the birth spin-down luminosity of the pulsar in Eq. \eqref{eq:L_t_fo_tau_0} and \eqref{eq:L_t_fo_tau_02} and by setting them equal 
\begin{equation}
 \left(\frac{\tau_c}{\tau_0} \right)^{\beta} = \left( 1+\frac{t_{\rm{age}}}{\tau_0} \right)^{\beta}
\end{equation}
we find
\begin{equation}
\label{eq:tau0_1234}
 \tau_0 = \tau_{\rm{c}} - t_{\rm{age}}.
\end{equation}
This equation is used in Section \ref{sec:Injec}. Therefore we choose $t_{\rm{age}}$ and $n$, calculate $\tau_{\rm{c}}$ and $L_{\rm{age}}$ using the measured value of $P$ and $\dot{P}$, calculate $\tau_0$ from Eq.~\eqref{eq:tau0_1234} and lastly $L_0$ from Eq.~\eqref{eq:L_t_fo_tau_02}. All parameters are now known and we can obtain $L(t)$ from Eq.~\eqref{eq:L_t_fo_tau_023}.

\end{document}